\documentclass[reprint,amsmath,amssymb,aps,prl,twocolumn,showpacs,superscriptaddress]{revtex4-1}
\usepackage[pdftex]{graphicx}
\usepackage{color}
\usepackage{amssymb}
\newcommand{\unit}[2]{\ensuremath{{#1}\,\mathrm{#2}}}

\begin{document}

\title{Ultrafast Electron Dynamics in the Topological Insulator Bi$_2$Se$_3$ Studied by Time-Resolved Photoemission Spectroscopy}

\author{J.~A. Sobota}
\author{S.-L. Yang}
\affiliation{Stanford Institute for Materials and Energy Sciences, SLAC National Accelerator Laboratory, 2575 Sand Hill Road, Menlo Park, CA 94025, USA}
\affiliation{Geballe Laboratory for Advanced Materials, Department of Applied Physics, Stanford University, Stanford, CA 94305, USA}
\affiliation{Department of Physics, Stanford University, Stanford, CA 94305, USA}
\author{D. Leuenberger}
\affiliation{Stanford Institute for Materials and Energy Sciences, SLAC National Accelerator Laboratory, 2575 Sand Hill Road, Menlo Park, CA 94025, USA}
\affiliation{Geballe Laboratory for Advanced Materials, Department of Applied Physics, Stanford University, Stanford, CA 94305, USA}
\author{A.~F. Kemper}
\affiliation{Lawrence Berkeley National Lab, 1 Cyclotron Road, Berkeley, CA 94720}
\author{J.~G. Analytis}
\affiliation{Department of Physics, University of California, Berkeley, California 94720, USA}
\author{I.~R. Fisher}
\affiliation{Stanford Institute for Materials and Energy Sciences, SLAC National Accelerator Laboratory, 2575 Sand Hill Road, Menlo Park, CA 94025, USA}
\affiliation{Geballe Laboratory for Advanced Materials, Department of Applied Physics, Stanford University, Stanford, CA 94305, USA}
\author{P.~S. Kirchmann}
\email{kirchman@stanford.edu}
\affiliation{Stanford Institute for Materials and Energy Sciences, SLAC National Accelerator Laboratory, 2575 Sand Hill Road, Menlo Park, CA 94025, USA}
\author{T.~P. Devereaux}
\affiliation{Stanford Institute for Materials and Energy Sciences, SLAC National Accelerator Laboratory, 2575 Sand Hill Road, Menlo Park, CA 94025, USA}
\affiliation{Geballe Laboratory for Advanced Materials, Department of Applied Physics, Stanford University, Stanford, CA 94305, USA}
\author{Z.-X. Shen}
\email{zxshen@stanford.edu}
\affiliation{Stanford Institute for Materials and Energy Sciences, SLAC National Accelerator Laboratory, 2575 Sand Hill Road, Menlo Park, CA 94025, USA}
\affiliation{Geballe Laboratory for Advanced Materials, Department of Applied Physics, Stanford University, Stanford, CA 94305, USA}
\affiliation{Department of Physics, Stanford University, Stanford, CA 94305, USA}

\date{\today}
\begin{abstract}
We characterize the topological insulator Bi$_2$Se$_3$ using time- and angle- resolved photoemission spectroscopy.  By employing two-photon photoemission, a complete picture of the unoccupied electronic structure from the Fermi level up to the vacuum level is obtained.  We demonstrate that the unoccupied states host a second, Dirac surface state which can be resonantly excited by 1.5~eV photons.  We then study the ultrafast relaxation processes following  optical excitation.  We find that they culminate in a persistent non-equilibrium population of the first Dirac surface state, which is maintained by a meta-stable population of the bulk conduction band.  Finally, we perform a temperature-dependent study of the electron-phonon scattering processes in the conduction band, and find the unexpected result that their rates decrease with increasing sample temperature.  We develop a model of phonon emission and absorption from a population of electrons, and show that this counter-intuitive trend is the natural consequence of fundamental electron-phonon scattering processes. This analysis serves as an important reminder that the decay rates extracted by time-resolved photoemission are not in general equal to single electron scattering rates, but include contributions from filling and emptying processes from a continuum of states. 
\end{abstract}
\maketitle

\section{Introduction}
Three dimensional topological insulators (TIs) are fascinating materials characterized by an insulating bulk electronic band structure with a metallic, conductive surface state (SS). The SS can be described by a Dirac equation for massless fermions, and is guaranteed to cross the band gap separating the bulk valence band (VB) and conduction band (CB) \cite{Fu2007,Zhang2009,Chen2009,Xia2009,Qi2011}. In addition, the SS is strongly spin-polarized with the electrons' spin-orientations locked perpendicular to their momenta \cite{Hsieh2009,Hsieh2009a}.  This novel spin texture results in suppressed backscattering \cite{Roushan2009} and makes the materials attractive for spintronics applications \cite{Garate2010,Pesin2012}. 

Because of these intriguing properties, TIs have been the subject of intense investigation using ultrafast time-resolved techniques.  These studies aim to unveil the fundamental scattering properties of electrons in TIs by resolving their dynamical processes in real time.  Ultrafast optical techniques monitor transient changes in the optical properties of a material, such as reflectivity.  These studies have yielded valuable insights on coherent phonon generation \cite{Qi2010,Kumar2011,Glinka2013}, spatial diffusion of photoexcited carriers \cite{Kumar2011}, and the timescales for electron-phonon scattering \cite{Qi2010,Hsieh2011a,Kumar2011,Glinka2013}.  A shortcoming of these techniques is that they lack momentum and energy resolution, and are limited in their ability to directly distinguish surface and bulk effects.  Time- and angle- resolved photoemission spectroscopy (trARPES) overcomes these limitations by adding momentum and energy resolution, allowing the material's response to be studied directly within its electronic band structure.  With these capabilities, trARPES has provided deeper understanding on issues such as bulk-to-surface scattering \cite{Sobota2012,Hajlaoui2012,Wang2012a,Hajlaoui2013}, electron-phonon coupling \cite{Sobota2012, Hajlaoui2012,Wang2012a,Crepaldi2012,Hajlaoui2013,Crepaldi2013}, unoccupied electronic structure \cite{Niesner2012,Sobota2013}, and novel topological states of matter \cite{Wang2013}.

In this article we demonstrate the wealth of information that can be gained about TIs from trARPES experiments.  In particular, we employ both one-photon photoemission spectroscopy (1PPE) and two-photon photoemission spectroscopy (2PPE) to study the electronic structure and dynamics of the prototypical TI Bi$_2$Se$_3$.  We first show that 2PPE resolves the unoccupied electronic structure with unprecedented clarity. This allows us to unambiguously identify a second SS well above the Fermi level $E_F$.  We find that optical excitation of $n$-type Bi$_2$Se$_3$ with 1.5~eV photons drives a direct optical transition precisely into this state, thereby laying the foundation for direct ultrafast optical coupling to a topological SS \cite{Sobota2013}. 

We continue by reporting the electron dynamics following this 1.5~eV excitation using time-resolved 2PPE. For this study we utilize $p$-type Bi$_2$Se$_3$, which has a completely unoccupied SS in equilibrium, thereby granting us a clear view of the electron relaxation through the SS.  We resolve the rapid decay of electrons to lower energy states via inter- and intra-band phonon-mediated scattering processes.  Within 2~ps a meta-stable population forms at the CB edge.  This population acts as an electron reservoir which fills the SS with a steady supply of carriers for \unit{\sim10}{ps}. This persistent occupation of a metallic state is a unique situation, since the absence of a bandgap in metallic bands typically means that there is no barrier to rapid recombination. This long-lived spin-textured metallic population may present a channel in which to drive transient spin-polarized currents \cite{Sobota2012}.

Finally, we examine the role of electron-phonon scattering processes by studying the dynamics of the CB as a function of sample temperature.  We discover decreasing population relaxation rates with increasing sample temperature.  As higher temperatures correspond to higher phonon occupations, this result might seem counter-intuitive at first glance.  However, we show that this behavior is reproduced by a straightforward electron-phonon scattering model.  While we aim to achieve only qualitative agreement in this work, this model has the potential to extract detailed, quantitative information about electron-phonon coupling from trARPES data.

\section{Methods}

\begin{figure}
\resizebox{\columnwidth}{!}{\includegraphics{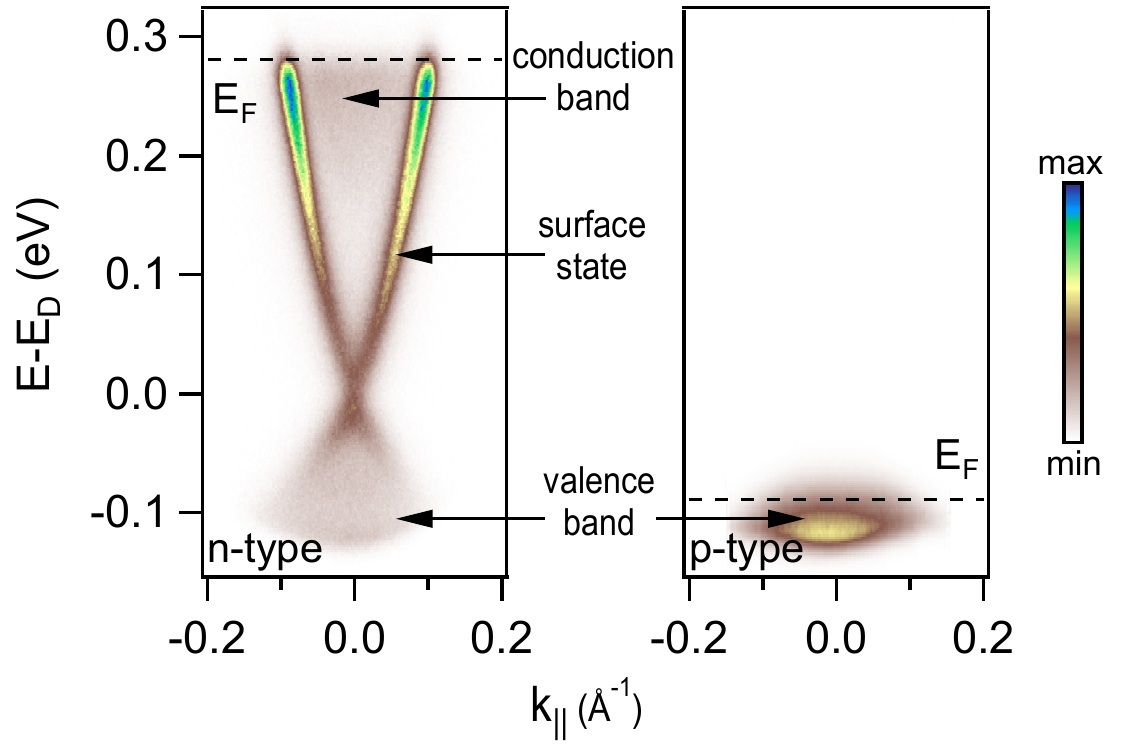}}
\caption{The band structure of the topological insulator Bi$_2$Se$_3$ as measured by ARPES. The band filling is determined by the bulk doping level. In an $n$-type sample (left), the conduction band, surface state, and valence band are occupied.  In a $p$-type sample (right), only the valence band is occupied.
\label{TI_ARPES}}
\end{figure}

Single crystals of Bi$_2$Se$_3$ were synthesized as described in Ref. \onlinecite{Sobota2012}.  $p$-type crystals were achieved by Mg-doping.  Fig. \ref{TI_ARPES} shows typical ARPES spectra for the samples used in this work.  The difference in the doping levels of $n-$ and $p-$type samples is clearly exhibited by the energetic position of $E_F$ and thus the band filling.


\begin{figure}
\resizebox{\columnwidth}{!}{\includegraphics{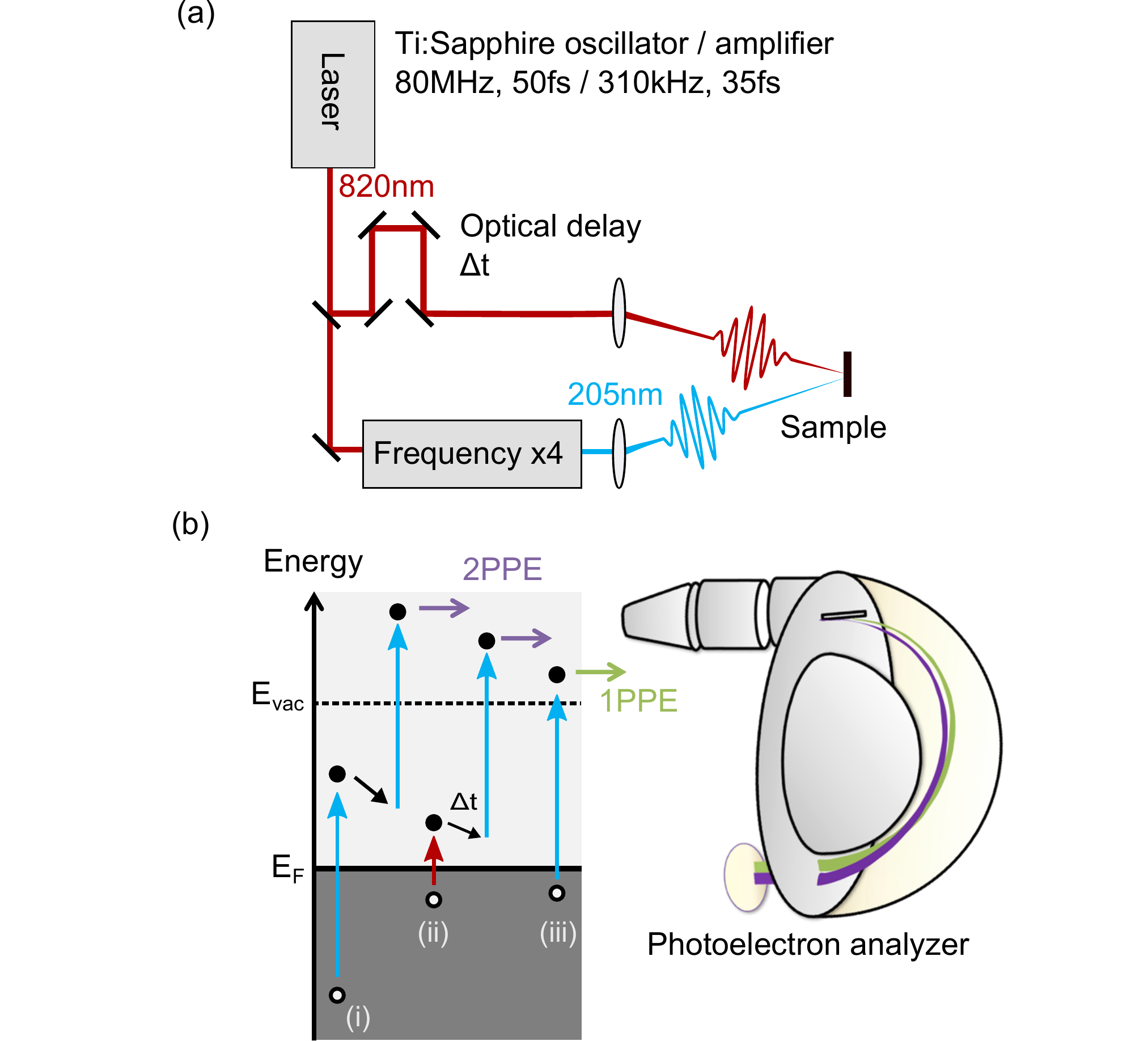}}
\caption{(a) The experimental setup used for the pump-probe photoemission experiments described in this work.  A laser generates ultrafast laser pulses at 820~nm (1.5~eV). A portion of the beam is used to optically excite the sample, and the remaining beam is frequency quadrupled to 205~nm (6.0~eV) to probe by photoemission. A variable path length allows for a tunable delay between the two optical paths.  (b) Schematic of the photoemission processes utilized in this work. (i) Monochromatic 2PPE using 6~eV photons, (ii) Bichromatic 2PPE using time-delayed 1.5~eV and 6~eV photons, and (iii) 1PPE using 6~eV photons.  The photoelectrons are collected by a hemispherical photoelectron analyzer.
\label{ExpSetup}}
\end{figure}

We illustrate our setup for trARPES in Fig. \ref{ExpSetup}(a). A Ti-Sapphire oscillator operating at \unit{80}{MHz}, or regenerative amplifier at 310~kHz, generates 820~nm (1.5~eV) infrared laser pulses which are split into two paths. In one path the pulse frequency is quadrupled to produce a 205~nm (6.0~eV) ultraviolet  beam.  The other path includes a delay stage which varies the optical path length.  Both beams are focused on the sample in an ultrahigh vacuum chamber.  By tuning the delay stage, the temporal delay between both pulses can be varied. 

This setup permits three primary modes of measurement, as summarized in Fig. \ref{ExpSetup}(b).  Process (i) represents 2PPE performed with the 6~eV pulses.  The entire two-photon process occurs within the pulse duration of the incident 6~eV pulse.  Since no time delay is introduced, this is a static measurement.   However, due to the finite time duration of the laser pulses (of order \unit{100}{fs} in our case), electron relaxation processes can occur in the intermediate states before the electron is photoemitted.  As will be shown below, this mode of measurement grants access to the unoccupied electronic structure between $E_F$ and the vacuum level $E_\textrm{vac}$ \cite{Haight1995,Petek1998,Weinelt2002}.

Process (ii) represents time-resolved 2PPE performed with time-delayed 1.5 and 6~eV pulses.  Here the 1.5~eV photon plays the role of a pump by optically exciting electrons within the material's band structure, while the 6~eV photon serves to probe the transiently modified electron distribution by photoemission. A complete movie of the excitation and relaxation dynamics is obtained by measuring the photoemission spectrum as a function of pump-probe delay \cite{Haight1995,Petek1998,Weinelt2002}.

Finally, process (iii) represents 1PPE performed with 6~eV pulses.  This is equivalent to conventional ARPES, and permits measurement of the equilibrium, occupied electronic structure \cite{Hufner1995}. 

We note the simplicity and flexibility of the setup which is achieved by using the same 6~eV photon energy for both 1PPE and 2PPE processes.  Our experimental configuration is identical for all measurements, except for the incident photon intensities.  Since 1PPE scales linearly with peak light intensity and 2PPE scales quadratically, we switch between 1PPE and 2PPE acquisition modes merely by tuning the intensity.  This is a significant distinction between our measurement and standard 2PPE measurements: Conventional 2PPE is performed with the photon energies chosen to be less than the sample work function, so that an overwhelmingly intense 1PPE signal is avoided. Here we deliberately use a photon energy larger than the work function, allowing us to perform both 1PPE and 2PPE with the same 6~eV pulses.  To avoid the intense 1PPE signal while measuring in mode (i), we operate the analyzer such that only electrons with energy $>E_F$ are collected

All measurements are performed in an ultrahigh vacuum chamber with pressure maintained below 1$\times$10$^{-10}$ Torr to minimize sample aging.  The emission angle and kinetic energy of the photoelectrons are resolved by a hemispherical electron analyzer.  The setup has a total energy resolution of \unit{<22}{meV} (limited by the bandwidth of the 6eV pulses) and angular resolution $<0.5^\circ$.  The temporal resolution, defined by the cross-correlation of the 1.5 and 6~eV pulses, is about 160~fs.

The density functional theory calculations were done with the PBE exchange functional \cite{Perdew1996} using the full potential (linearized) augmented plane-wave method as implemented in the \textsc{wien2k} package\cite{Blaha2001}. The calculations were based on the experimentally determined crystal structure reported in Ref. \onlinecite{Nakajima1963}.  The Bi$_2$Se$_3$ slab consists of 6 quintuple layers, separated by a vacuum of  $\sim 25$~\AA. We used a $k$-mesh size of $15\times15\times1$, and utilized spin-orbit coupling in the self-consistent calculations unless otherwise indicated.

\section{Results \& Discussion}
\subsection{Optical coupling to a second Dirac surface state}

\begin{figure}
\resizebox{\columnwidth}{!}{\includegraphics{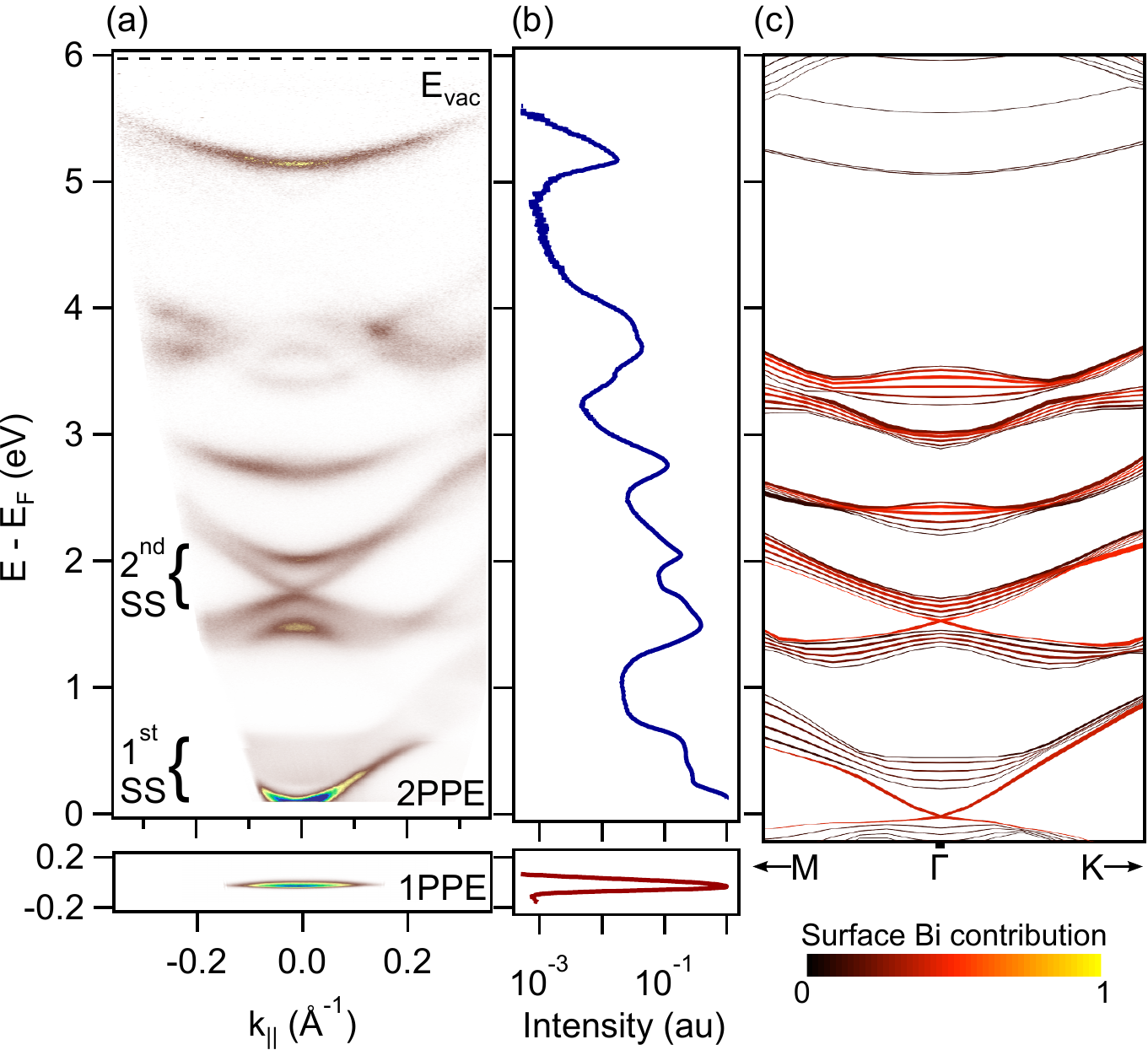}}
\caption{(a) 2PPE and 1PPE spectra of $p$-type Bi$_2$Se$_3$. The 1PPE spectrum contains only a narrow band of intensity from the occupied portion of the VB above the low-energy cutoff. The 2PPE spectrum reveals states from $E_F$ up to $E_{\textrm{vac}}$, including the well-known Dirac state around $\sim0.5$~eV, as well as a second Dirac state around $\sim1.8$~eV. The intensities are rescaled exponentially as a function of energy to make all features visible on the same image. (b) Momentum-integrated energy distribution curves showing the dynamic range of the 2PPE intensity. (c) Slab band structure calculation, which shows a one-to-one correspondence with the observed features.  Figure reproduced from Ref. \onlinecite{Sobota2013} (Copyright 2013 by the American Physical Society).
\label{2PPE_Overview}}
\end{figure}

We begin by presenting the 2PPE and 1PPE spectra of $p$-type Bi$_2$Se$_3$ in the upper and lower panels of Fig. \ref{2PPE_Overview}(a).  These spectra correspond to processes (i) and (iii) of Fig. \ref{ExpSetup}(b). The 1PPE spectrum consists of only a narrow band of intensity from the occupied portion of the VB above the low-energy cutoff, as in Fig. \ref{TI_ARPES} \cite{Sobota2012}.  The 2PPE spectrum, on the other hand, reveals states all the way from $E_F$ up to $E_{\textrm{vac}}$.  Just above $E_F$ we find the familiar CB and SS, which would be occupied in an $n$-type sample.   Note that the spectral intensity spans a dynamic range of $\sim 3$ orders of magnitude (see Fig. \ref{2PPE_Overview}(b)) and so the intensities of \ref{2PPE_Overview}(a) are rescaled exponentially with energy to allow all features to be displayed simultaneously.

Before discussing these results in greater detail, it is important to understand the origin of the observed spectral features.  In principle, the 2PPE spectral intensity can be complicated by contributions from both initial and final states, in addition to the intermediate states which are the subject of our study \cite{Petek1998,Weinelt2002}.  To check whether this is the case, we show the calculated band structure in Fig. \ref{2PPE_Overview}(c). While the energy scales are not reproduced exactly, there is an unambiguous one-to-one correspondence between the observed and calculated features. This demonstrates that the 2PPE spectrum is dominated by the intermediate states. 

It is not immediately clear why a strong dependence of the initial states is not observed.  This is likely due to the fact that the electronic states populated by the 6~eV photons are very high in energy, which means there is a large phase space for decay, with correspondingly short lifetimes ($\sim$1~fs) \cite{Giu05}.  We therefore expect significant inter-and intra- band scattering to occur within the duration of the 6~eV pulse.  In effect, these relaxation processes rapidly redistribute electrons among all the unoccupied bands.  

Now that we understand the origin of the 2PPE spectral features, we proceed by discussing the electronic states located between $1.3$ and \unit{2.4}{eV} of Fig. \ref{2PPE_Overview}(a).  The band structure here is strikingly similar to the VB, CB, and SS bands near $E_F$. In fact, Niesner \emph{et. al.} first identified the linearly dispersive band as an unoccupied topologically protected, spin chiral SS, while the upper and lower bands are the corresponding bulk bands\cite{Niesner2012}. To distinguish these two sets of bands, we refer to the Dirac cone near $E_F$ as the $1^\textrm{st}$ SS, and the higher energy Dirac cone as the $2^\textrm{nd}$ SS. The corresponding bulk bands are referred to as the $1^\textrm{st}$ and $2^\textrm{nd}$ VB and CB. The $2^\textrm{nd}$ SS disperses with a velocity of  \unit{3.3\times10^5}{m/s} at the Dirac point, which is comparable to the velocity of \unit{5.4\times10^5}{m/s} typically measured for the Fermi velocity of the $1^\textrm{st}$ SS \cite{Kuroda2010}.  Calculations confirm the assignment of these bands to bulk and surface states \cite{Niesner2012, Eremeev2013, Sobota2013}. Moreover, just like the $1^\textrm{st}$ SS, the $2^\textrm{nd}$ SS exists only in the presence of crystal spin-orbit interaction \cite{Niesner2012, Eremeev2013, Sobota2013}. This suggests that both SSs share the same physical origin, as they both arise due to symmetry inversion of bulk states in the presence of strong spin-orbit coupling.

As we shall show, the existence of the $2^\textrm{nd}$ SS could be highly relevant to studies utilizing ultrafast optical excitation of TIs. These experiments typically use a photon energy of \unit{1.5}{eV} due to the use of a Ti:Sapphire laser source. Furthermore, $n$-type samples are more commonly used than $p$-type.  We have therefore investigated the effect of optically exciting $n$-type Bi$_2$Se$_3$ with 1.5~eV photons.

\begin{figure}
\resizebox{\columnwidth}{!}{\includegraphics{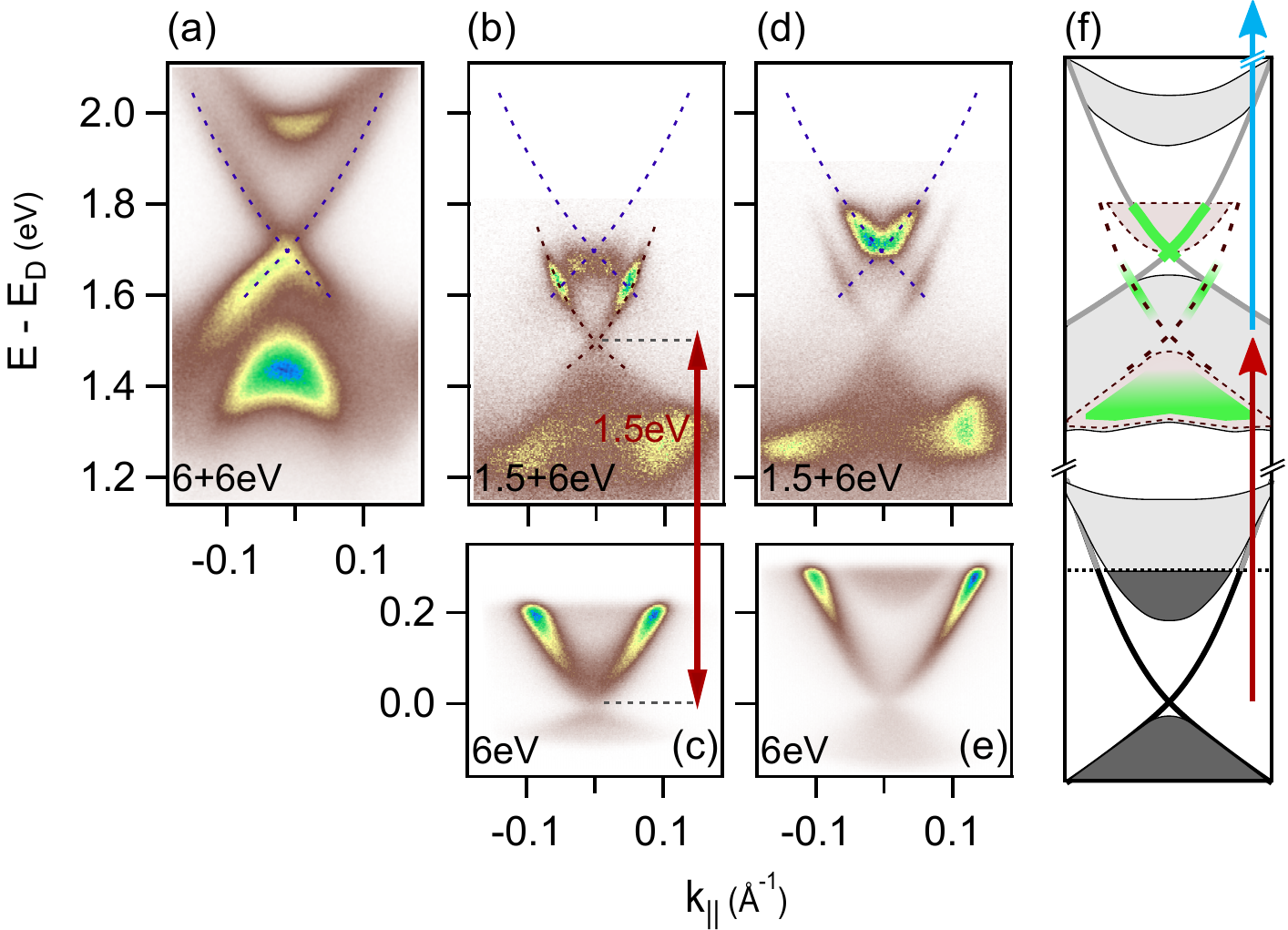}}
\caption{Optical excitation of the $2^\textrm{nd}$ Dirac cone using 1.5~eV photons on $n$-type Bi$_2$Se$_3$.  Dashed lines are guides to the eye.(a) For reference, a 6~eV 2PPE spectrum of the $2^\textrm{nd}$ Dirac cone is shown. (b) 2PPE spectrum using 1.5~eV photons to populate the $2^\textrm{nd}$ SS, and 6~eV photons to photoemit. (c) 6~eV 1PPE measurement of the occupied states.  Panels (d) and (e) are the same measurements, but for a sample with a higher level of $n$-type doping.  (f) Cartoon picture of the 1.5~eV excitation, 6~eV probe process.  The resulting 2PPE spectrum can be understood as a projection of the initial states onto the intermediate states.  Both resonant and non-resonant 2PPE processes are resolved.  Green shading represents the resonant processes.
\label{pump_2ndcone}}
\end{figure}

In Fig.~\ref{pump_2ndcone}(a) we first show for reference the unoccupied $2^\textrm{nd}$ SS obtained by \unit{6}{eV} 2PPE.  The effect of 1.5~eV excitation is shown in Fig.~\ref{pump_2ndcone}(b).  Specifically, we use 1.5~eV to optically excite the sample, and 6~eV to photoemit.  This corresponds to process (ii) of Fig. \ref{ExpSetup}(b), with the time delay set to zero.  The corresponding 1PPE spectrum of the occupied band structure is shown in Fig.~\ref{pump_2ndcone}(c).  A similar set of measurements, but for a sample with a slightly higher doping level, is shown in Figs.~\ref{pump_2ndcone}(d) and (e).  

The most striking feature in the spectra of Figs.~\ref{pump_2ndcone}(b) and (d) is a linearly dispersing feature.  At first glance, this would appear to represent the $2^\textrm{nd}$ SS, with features above and below representing the $2^\textrm{nd}$ CB and VB.  However, the appearance of these spectra is highly misleading:  Comparison with Fig.~\ref{pump_2ndcone}(a) shows that this cannot be the $2^\textrm{nd}$ SS, since the energies of the Dirac points do not match.  We instead identify the linearly dispersing feature as 2PPE from the $1^\textrm{st}$ SS.  This assignment is supported by the fact that its Dirac point is located precisely 1.5~eV above the Dirac point of the $1^\textrm{st}$ SS.  

In general, the spectra of Figs.~\ref{pump_2ndcone}(b) and (d) can be understood as a projection of the initial states (represented by (c) and (e)) onto the intermediate states 1.5~eV above (represented by (a)).  The features are brightest where the initial and intermediate states overlap in energy and momentum.  This is most dramatic in Fig.~\ref{pump_2ndcone}(b), where the projection of the upper half of the $1^\textrm{st}$ SS onto the lower half of the $2^\textrm{nd}$ SS results in two brights spots of intensity.  These are known as resonant 2PPE processes, since the 1.5~eV photon resonantly excites a state 1.5~eV above the initial state at the same wave vector.  The less intense features represent non-resonant 2PPE processes \cite{Petek1998,Weinelt2002}.  A cartoon summary is provided in Fig.~\ref{pump_2ndcone}(f), where resonant 2PPE processes are highlighted in green.

We note a significant difference between the 2PPE results of Figs. \ref{2PPE_Overview} and \ref{pump_2ndcone}.  For Fig. \ref{2PPE_Overview} we emphasized that the dispersion of the initial states plays a negligible role in the appearance of the 2PPE spectrum, while the results of Fig. \ref{pump_2ndcone} are a complicated combination of initial- and intermediate- state effects.  While the reason for this difference is not fully understood, we believe it is related to the fact that  1.5~eV photons promote electrons to much lower energy than 6~eV photons.  This means there is substantially less phase space for decay of the photo-excited electrons, and correspondingly longer electron lifetimes.  Therefore, the rapid scattering processes which are responsible for washing out the initial state effects for $6+6$~eV 2PPE are much less significant in the case of $1.5+6$~eV 2PPE.

\subsection{Persistent surface state population}

We now use time-resolved 2PPE to investigate the ultrafast dynamics of optically excited electrons in Bi$_2$Se$_3$.  We confine our attention to the $1^\textrm{st}$ SS, CB, and VB.  Full details of this experiment are reported in Ref. \onlinecite{Sobota2012}. For this study we utilize $p$-type samples, since the initially unoccupied SS provides for an exceptionally clear view of electron relaxation through the SS.

\begin{figure}
\resizebox{3.35in}{!}{\includegraphics{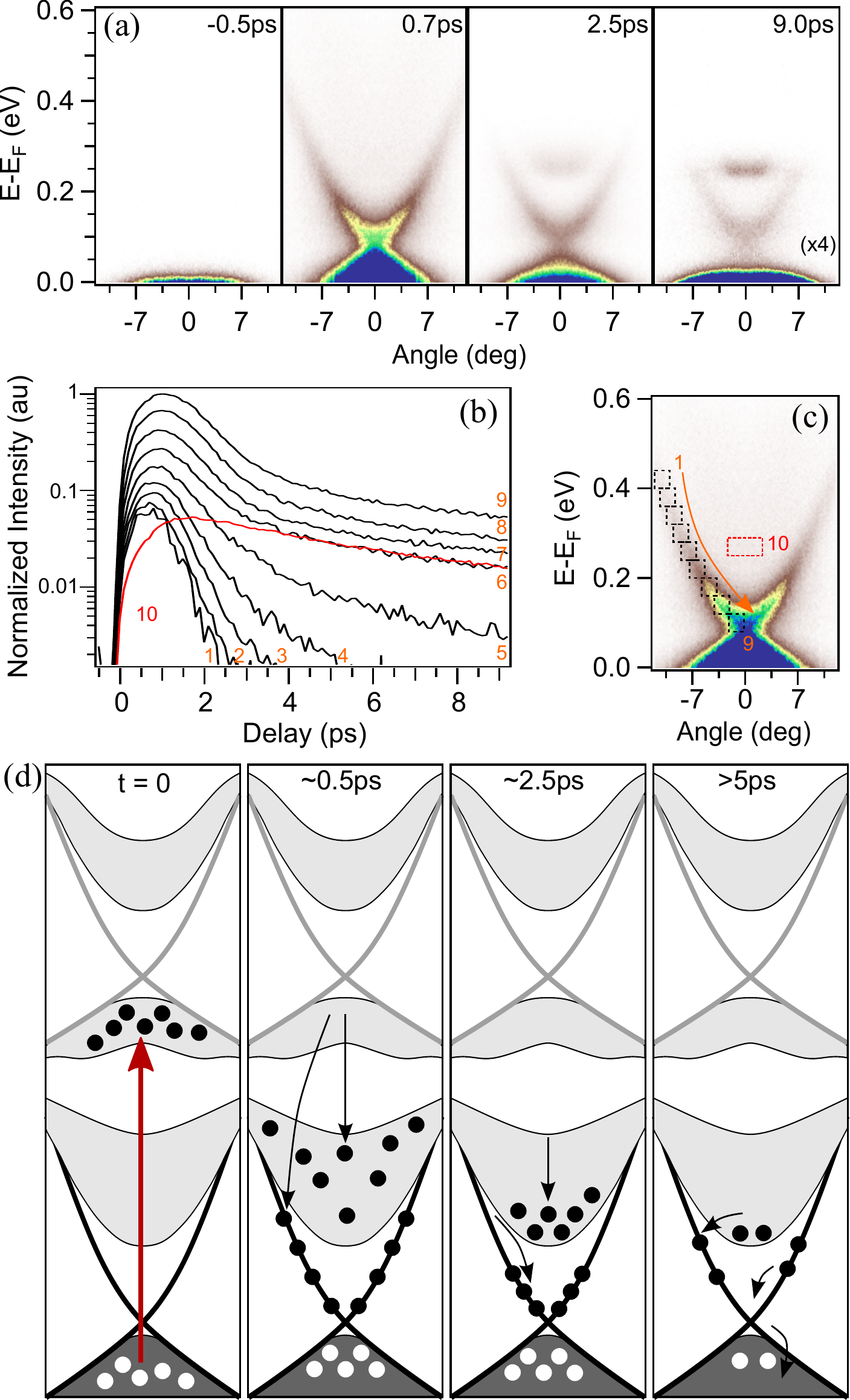}}
\caption{ (a) Overview of the dynamics generated by 1.5~eV optical excitation of $p$-type Bi$_2$Se$_3$.  The SS and CB are unoccupied in equilibrium, but are transiently populated and undergo subsequent decay processes.  The intensity in the fourth panel is multiplied by a factor of 4 to allow weak features to be seen.  (b) Transient photoemission intensities within the integration windows indicated in the subsequent panel (c).  Curve \#10 corresponds to the integration window over the CB edge, and is normalized to match the intensity of the SS intensities. The slow component of the SS decay has the same 5.95(2)~ps timescale as the CB population decay. (d)  Schematic of the transitions and scattering processes generated by 1.5~eV excitation of $p$-type Bi$_2$Se$_3$, including the direct optical transition, scattering into SS and CB, intra-band scattering of the CB and SS, and the CB-to-SS scattering responsible for a persistent SS population.  Figure adapted from Ref. \onlinecite{Sobota2012}.
\label{1p5_ptype}}
\end{figure}

An overview of the dynamics upon 1.5~eV excitation with 26~$\mu$J/cm$^2$ absorbed fluence is shown in Fig. \ref{1p5_ptype}(a): Starting in equilibrium ($t<0$), the SS and CB are initially unoccupied.  Upon excitation, high-lying electronic states like those in Fig. \ref{pump_2ndcone} are populated. After $\sim$0.7~ps those electrons scatter to lower energy, populating the SS and CB. After $\sim$2.5~ps the SS and CB populations have significantly decayed and energetically relaxed towards the bottom of their respective bands. The subsequent dynamics is much slower and persists for \unit{>9}{ps}. Obstructed from further decay due to the bandgap, the CB electrons form a meta-stable population at the CB edge.  Intriguingly, this CB population is accompanied by a persistent population in the SS, but only energetically below the CB edge.

The observation of a long-lived CB population is not unusual in semiconductors \cite{Sze2006}.  The longevity of this population is due to the fact that the electron-phonon scattering processes responsible for energy relaxation within the band are incapable of recombining electrons with the VB.  This is because the highest phonon energy in Bi$_2$Se$_3$ is $\approx$\unit{23}{meV} \cite{Richter1977}, while the bandgap is \unit{200}{meV} \cite{Chen2010}.  The simultaneous persistence of the metallic SS population, however, is surprising since the absence of a bandgap means there is no barrier to rapid recombination. We attribute this SS population to a continuous filling from the meta-stable CB population.  Evidence for this can be found in the population dynamics.  In Fig.~\ref{1p5_ptype}(b) we plot the transient photoemission intensity obtained by integrating within the windows indicated in Fig.~\ref{1p5_ptype}(c).  Above the CB, the SS decays with a single exponential.  However, below the CB a second slower component is observed, reflecting the presence of an additional filling channel.  Moreover, the decay rate of this slower component matches that of the population decay at the CB edge (see curves \#6 and \#10).  The agreement of these decay rates is direct evidence that the persistent SS population is due to continuous filling from the meta-stable CB population.  A cartoon summarizing these processes is shown in Fig. \ref{1p5_ptype}(d).  We note that similar bulk-to-surface scattering behavior has been observed in Si, which however lacks a metallic SS crossing $E_F$ \cite{Weinelt2004,Tanaka2009}. More recently, trARPES experiments on $n$-type Bi$_2$Se$_3$ have also identified signatures of a bulk-to-surface scattering channel mediated by phonons \cite{Hajlaoui2012,Wang2012a,Hajlaoui2013}.

An exponential fit to the CB population decay (curve \#10 in Fig.~\ref{1p5_ptype}(b)) gives a population lifetime of \unit{5.95(2)}{ps}. The decay mechanisms responsible for this lifetime may have several contributions.  We have argued that electron-phonon coupling cannot be responsible for recombination due to the low phonon energies.  Direct recombination via photon emission is a possibility, though the timescale for such a process is typically \unit{\gg 1}{ns}, and thus seems unlikely to be relevant on our measurement timescale \cite{Schroder2005}. The CB decay could also be attributed to spatial diffusion of CB electrons away from the probed volume of the sample. A characteristic timescale for this process was estimated to be $\sim 21$~ps\cite{Sobota2012}.  Of course, the inter-band scattering channel into the SS provides another effective recombination mechanism.  Future studies utilizing Bi$_2$Se$_3$ thin films to mitigate spatial diffusion  may be helpful in disentangling these processes.

\subsection{Energy relaxation via electron-phonon scattering in the conduction band}

\begin{figure}
\resizebox{\columnwidth}{!}{\includegraphics{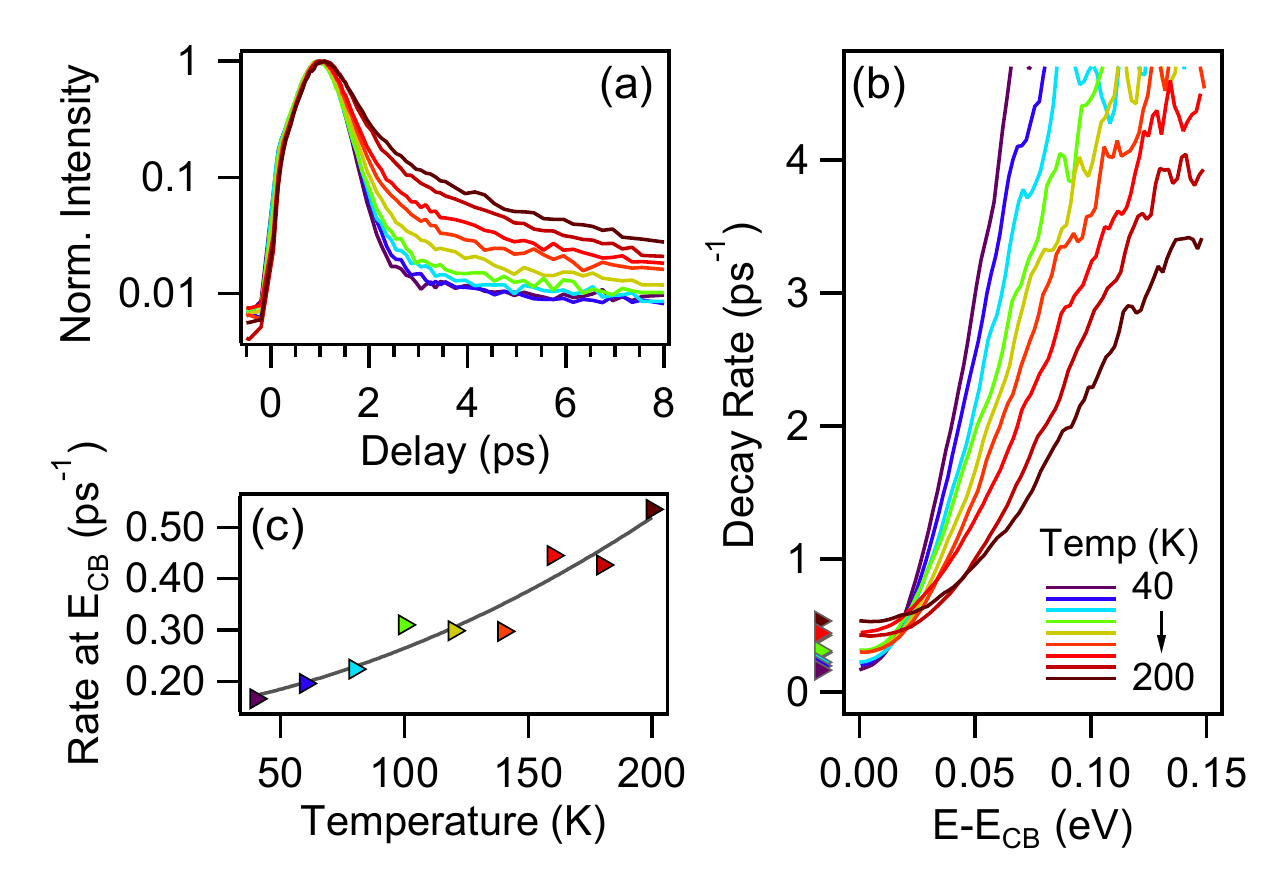}}
\caption{(a) Transient photoemission intensity at a fixed energy $E-E_{\textrm{CB}} = $0.08~eV for temperatures from 40 to 200~K in 20~K steps. The population relaxation becomes slower for increasing temperature. (b) The extracted population decay rates as a function of temperature and energy.  The rates at the CB edge ($E=E_{\textrm{CB}}$) are marked with triangles and plotted as a function of temperature in (c). The smooth curve is a quadratic fit to the data points. 
\label{TempData}}
\end{figure}
We have argued that the optically excited electron population in the CB relaxes via electron-phonon scattering processes. Since the phonon population changes with sample temperature, we expect the relaxation rates to exhibit a temperature dependence. Accordingly, we have performed temperature-dependent studies of the electron relaxation in the CB of $p$-type Bi$_2$Se$_3$. 

Fig.~\ref{TempData}(a) shows the transient photoemission intensity at a fixed energy relative to the band edge $E_\textrm{CB}$  for  temperatures from 40 to 200~K in 20~K steps using an absorbed pump fluence of 1.3~$\mu$J/cm$^2$.  We utilize this low fluence to minimize sample heating and thus have a well defined sample temperature.  While the initial rise of the population does not exhibit any pronounced temperature dependence, the subsequent decay slows down with increasing temperature. We fit a single exponential to the initial decay and plot the resulting decay rates as function of energy in Fig.~\ref{TempData}(b).  There are two distinct temperature dependencies: near $E_\textrm{CB}$ the rates increase with increasing temperature, and at higher energies the rates decrease with increasing temperature.  To understand the origin of these two behaviors,  we must distinguish between two types of scattering processes: intra-band scattering, which conserves the electron number in the band, and band-emptying processes.  The decay rate at $E=E_{\textrm{CB}}$ cannot represent intra-band scattering, since the phase space for decay at the band edge is strictly zero.  Instead, the decay rate at $E_{\textrm{CB}}$ must represent a band-emptying rate.  The emptying rate extracted from the data is plotted as a function of temperature in Fig.~\ref{TempData}(c).  

We proceed to discuss the intra-band scattering processes that dominate away from $E_\textrm{CB}$. The dramatic decrease of these rates with increasing sample temperature seems counter-intuitive when one considers that the phonon population $n$ increases with increasing temperature $T$ for all phonon energies $\omega$, as represented by the Bose-Einstein distribution: \cite{Grimvall1981}

\begin{align}
n(\omega) = \frac{1}{e^{\omega/k_\textrm{B}T}-1}
\end{align}

This poses the question of what mechanism can result in decreasing rates as the temperature is increased. 


To address this question we model the change in electron distribution due to absorption and emission of phonons.  The corresponding scattering rates follow from Fermi's golden rule \cite{Grimvall1981}. For simplicity, we neglect any momentum dependence in the following discussion.  Under these assumptions, the rate for an electron to scatter out of the state at $\epsilon$ due to coupling with a phonon of energy $\omega$ is given by:

\begin{align}
	\Gamma_\textrm{abs}(\epsilon,\omega) & = \frac{2\pi}{\hbar} \lvert G(\omega) \vert^2 n(\omega) N(\epsilon + \omega) (1 - f(\epsilon + \omega)) \label{RateValues1} \\ 
	\Gamma_\textrm{emis}(\epsilon,\omega) & =\frac{2\pi}{\hbar} \lvert G(\omega) \vert^2 (1+n(\omega)) N(\epsilon - \omega) (1 - f(\epsilon - \omega))
	\label{RateValues2}
\end{align}

Here $\Gamma_\textrm{abs}$ ($\Gamma_\textrm{emis}$) denotes the rate attributable to phonon absorption (emission),  $N(\epsilon)$ is the electronic density of states (DOS) and  $f(\epsilon)$ is the electron occupation, which is a number between 0 and 1.  $G(\omega)$ is the matrix element describing the scattering process.  Physically, it encodes the electron coupling strength to phonons of energy $\omega$.

\begin{figure}
\resizebox{\columnwidth}{!}{\includegraphics{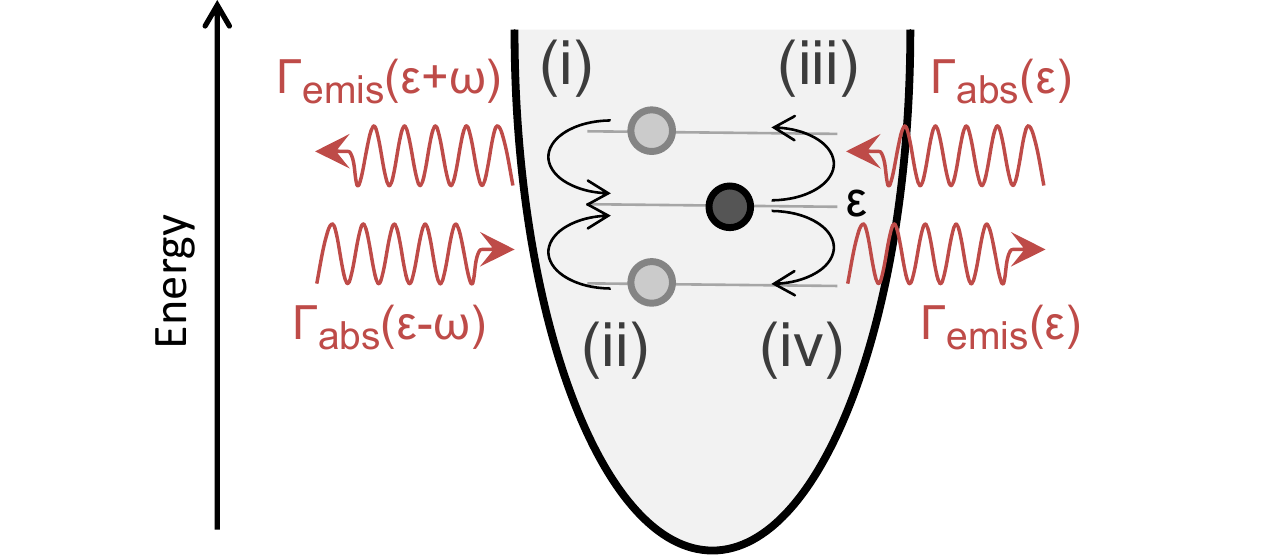}}
\caption{Cartoon of the four electron-phonon scattering processes relevant for the relaxation of electrons in the CB. (i) and (ii) represent the filling of the state at energy $\epsilon$ by phonon emission and absorption, respectively.  (iii) and (iv) represent the emptying of the state at energy $\epsilon$ by phonon absorption and emission, respectively.
\label{eph_cartoon}}
\end{figure}

In addition to scattering out of the state at $\epsilon$, we must also consider scattering in from other states.  There are therefore four processes which must be included, as represented in Fig.~\ref{eph_cartoon}.  We introduce the phonon DOS $F(\omega)$ to integrate over all $\omega$, and obtain rate equations describing the time evolution of $f(\epsilon)$:

\begin{equation}\begin{aligned}
	N(\epsilon)\frac{\partial f_\textrm{out}(\epsilon)}{\partial t} &= \int d\omega \, F(\omega)  [ \Gamma_\textrm{abs}(\epsilon,\omega)N(\epsilon) f(\epsilon)   
	\\ &+ \Gamma_\textrm{emis}(\epsilon,\omega)N(\epsilon) f(\epsilon)   ]  
\end{aligned}\label{ModelEq1}\end{equation}

\begin{equation}
\begin{aligned}
	N(\epsilon)\frac{\partial f_\textrm{in}(\epsilon)}{\partial t}   &=  \int d\omega \, F(\omega) [ \Gamma_\textrm{abs}(\epsilon-\omega,\omega) N(\epsilon-\omega)f(\epsilon-\omega)
		 \\ & +\Gamma_\textrm{emis}(\epsilon+\omega,\omega) N(\epsilon+\omega)f(\epsilon+\omega) ]
\end{aligned}\label{ModelEq2}
\end{equation} 

The complete dynamics of the model are then described by:

\begin{equation}
\frac{\partial f(\epsilon)}{\partial t} = \frac{\partial f_\textrm{in}(\epsilon)}{\partial t} - \frac{\partial f_\textrm{out}(\epsilon)}{\partial t} - \Gamma_\textrm{esc} f(\epsilon)
\end{equation}

The last term $\Gamma_\textrm{esc}$ represents the rate at which electrons escape the band.  This term can be set to zero to limit the calculation to intra-band processes only and conserve the electron number in the band.  

For the numerical calculation we shall assume $\left( 1- f(\epsilon)\right) \approx 1$ in Eqs. \ref{RateValues1} and \ref{RateValues2}. This is not required for the calculation, but is a reasonable assumption for the CB of $p$-type Bi$_2$Se$_3$ since it is completely unoccupied in equilibrium, and only perturbatively populated by the optical excitation.  The lattice temperature enters through the temperature dependence of the Bose-Einstein distribution $n(\omega)$.  In this formulation we assume that the lattice temperature remains constant during the electron relaxation process.  This assumption is consistent with experiments in which the lattice undergoes a much smaller transient change in  temperature as compared to the electrons due to its significantly higher specific heat \cite{Lisowski2004,Bovensiepen2007}.

Four quantities need to be specified to proceed with the calculation: $N(\epsilon)$, $\lvert G(\omega) \rvert^2$,  $F(\omega)$, and $f(\epsilon)\vert_{t=0}$, which defines the initial condition for the evolution of the electron distribution.  

\begin{figure}
\resizebox{\columnwidth}{!}{\includegraphics{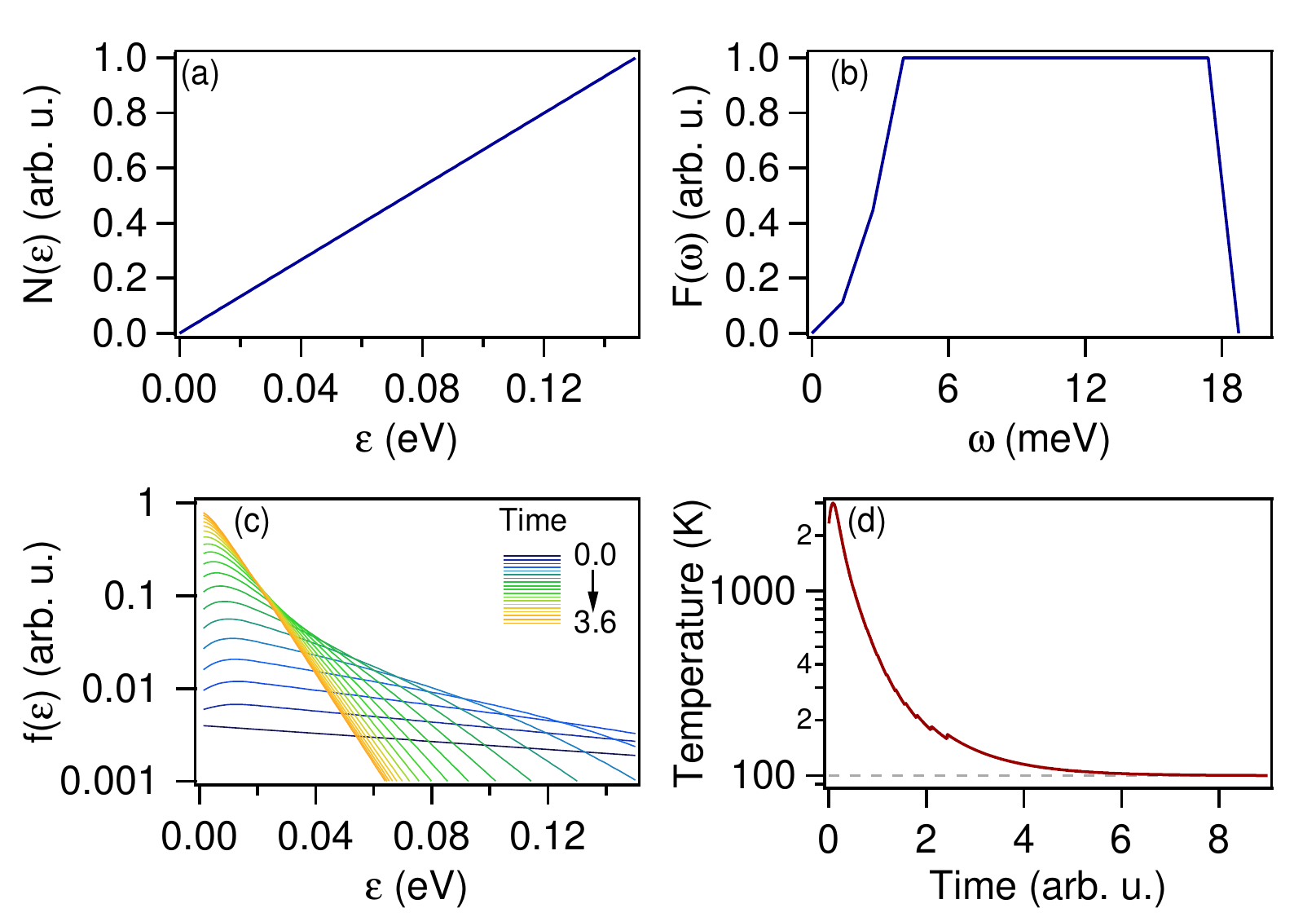}}
\caption{Results of the electron-phonon scattering model computed with a phonon temperature of 100~K and $\Gamma_\textrm{esc} = 0$.  (a) The electronic density of states and (b) the phonon density of states used as inputs for the calculation.  (c)  The computed electronic distribution as a function of time.  (d)  The extracted transient electronic temperature.
\label{SimDemo}}
\end{figure}

For $N(\epsilon)$ we assume a linearly increasing DOS, as shown in Fig. \ref{SimDemo}(a).  We base this assumption on a comparison to the DOS measured by scanning tunneling experiments \cite{Cheng2010}. Note that we only specify $N(\epsilon)$ up to a proportionality factor.  This implies that the temporal units of our calculation will be arbitrary.  Nevertheless, this will be sufficient to understand the qualitative behavior. The choice of  $\lvert G(\omega) \rvert^2$ is less straightforward.  We found that the choice of $\lvert G(\omega) \rvert^2$ does not qualitatively change the results.  To avoid detailed calculations of electron-phonon coupling matrix elements,  we shall assume that all phonon modes couple equally well to all electrons, so that $\lvert G(\omega) \rvert^2$  is simply a constant.  The $F(\omega)$ used for our modeling is shown in Fig. \ref{SimDemo}(b).  This choice is an approximation to the measured phonon DOS \cite{Rauh1981}.  Finally, we can use our trARPES data to make a suitable choice for $f(\epsilon)\vert_{t=0}$.  We use $f(\epsilon)\vert_{t=0} \propto e^{-\epsilon/\epsilon_0}$ with $\epsilon_0$ = 0.2~eV.

With these four assumptions, it is straightforward to compute the temporal evolution of $f(\epsilon)$ for any $T$ and $\Gamma_\textrm{esc}$.  The results for $T=100$~K with $\Gamma_\textrm{esc}=0$ are shown in Fig. \ref{SimDemo}(c).  Interestingly, the electronic distribution remains essentially exponential as it evolves, and eventually saturates to a steady-state solution.  In Fig. \ref{SimDemo}(d) we extract a transient electronic temperature by fitting $f(\epsilon)$ to an exponential distribution at each time.  Reassuringly, the electronic temperature settles at 100~K.  This demonstrates that this simple model captures the process by which electrons reach thermal equilibrium with the lattice.  In fact, it can be shown analytically that 	$\frac{\partial f_\textrm{in}}{\partial t} = 	\frac{\partial f_\textrm{out}}{\partial t}$ for $f \propto e^{-\epsilon/T}$, where $T$ is the lattice temperature.  In other words, the steady-state electronic distribution is a Boltzmann function at thermal equilibrium with the lattice.  (Note that Fermi-Dirac statistics need not be considered here due to the assumption $f\ll1$ made in this analysis).

\begin{figure}
\resizebox{\columnwidth}{!}{\includegraphics{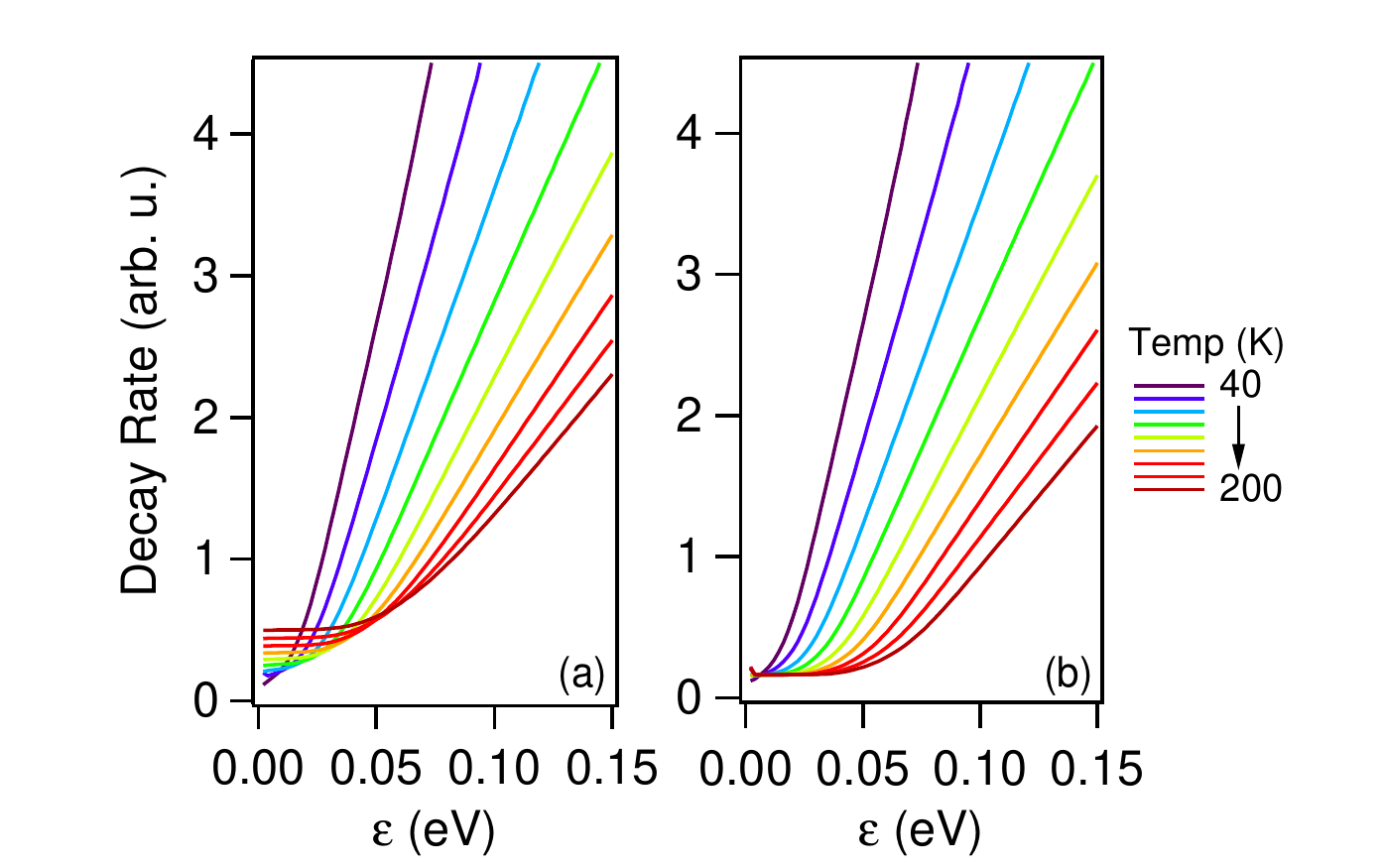}}
\caption{Temperature dependence of the electron-phonon scattering model.  (a)  The population decay rates for varying sample temperature.   $\Gamma_\textrm{esc}$ is set to the temperature-dependent values extracted from the fit in Fig. \ref{TempData}(c).  (b) The population decay rates, but with $\Gamma_\textrm{esc}$ set to the temperature-independent value of 0.17. }
\label{ModelTempDep}
\end{figure}

Now that we understand the behavior of this model for fixed $T$, we continue by varying the temperature.  Because the behavior of the rate curves of Fig. \ref{TempData}(b)  are strongly affected by the temperature-dependence of the emptying rate at $E_\textrm{CB}$, we set $\Gamma_\textrm{esc}$ equal to the fitted values in Fig. \ref{TempData}(c).  In Fig. \ref{ModelTempDep}(a) we show the extracted exponential decay rate as a function of $\epsilon$ and $T$.  The result agrees remarkably well with the experimental data of Fig. \ref{TempData}(b).  In particular, away from the band edge it exhibits a monotonic decrease of relaxation rate with increasing $T$.  In addition, the model reproduces the flattening of the rate curve at the value of $\Gamma_\textrm{esc}$ near the band edge.  

While the $T$-dependent $\Gamma_\textrm{esc}$ is required to reproduce the behavior of the experimental data near $E_\textrm{CB}$, we find it instructive to repeat the calculation for a fixed, $T$-independent value of $\Gamma_\textrm{esc}$.  The resulting rates are shown in Fig. \ref{ModelTempDep}(b).  With the $T$-dependence of $\Gamma_\textrm{esc}$ removed, only the temperature dependence of the intra-band electron-phonon scattering processes remains.  This clearly demonstrates that the counter-intuitive trend of increasing decay rates with decreasing $T$ can be attributed to fundamental electron-phonon scattering processes alone.  

\section{Conclusions \& Outlook}
In summary, we have characterized the electronic structure of Bi$_2$Se$_3$ with a combination of 1PPE and 2PPE techniques, performed time-resolved measurements on $p$-type Bi$_2$Se$_3$, and elucidated the rich electron dynamics generated by optical excitation.  In particular, we demonstrated that 2PPE with \unit{6}{eV} photons illuminates the states between $E_F$ and $E_{\textrm{vac}}$. This technique reveals a $2^\textrm{nd}$ Dirac SS located \unit{1.5}{eV} above the $1^\textrm{st}$ CB edge, and we showed how excitation with \unit{1.5}{eV} photons drives a direct optical transition between these two states.  

To our knowledge, this is the first measurement on any material  using a single photon energy for both 1PPE and 2PPE. This is achieved by using a photon energy exceeding the sample work function, which has conventionally been avoided in 2PPE measurements. Since the only experimental parameter  tuned between 1PPE and 2PPE modes of measurement is the intensity, this represents an exceptionally simple method to  determine both the occupied and unoccupied band structure of a material. This has exciting implications, for example, for materials such as the high-$T_c$ cuprates, where a measurement of both the occupied and unoccupied sides of the energy gap would provide insight on the particle-hole asymmetry of the superconducting and pseudogap states \cite{Hashimoto2010,Moritz2011}.

The demonstrated ability to optically excite the  $2^\textrm{nd}$ SS has a number of interesting implications.  The $2^\textrm{nd}$ SS is expected to have all the novel topological and spin properties which characterize the well-known $1^\textrm{st}$ Dirac SS of TIs \cite{Niesner2012,Eremeev2013}.  The results of Fig.  \ref{pump_2ndcone} demonstrate that it is resonantly populated by 1.5~eV photons.  This could have relevance for the number of existing studies on $n$-type TIs which have utilized 1.5~eV photons \cite{Qi2010,Kumar2011,Hsieh2011a,Wang2012a,Hajlaoui2012,Crepaldi2012,McIver2012}. These experiments have been interpreted without knowledge of the unoccupied topological SS, and it would be interesting to evaluate whether the transition into this state plays a role in the physics discussed in those results. Finally, the fact that it can be accessed by 1.5~eV photons is particularly appropriate for applications, since this is the fundamental photon energy provided by commercial ultrafast Ti:Sapphire lasers. This discovery therefore demonstrates a unique opportunity for direct ultrafast optical coupling to TI SSs. 

We then proceeded to study the electron dynamics initiated by optical excitation of $p$-type Bi$_2$Se$_3$.  We found that the dynamics culminate in a persistent non-equilibrium population of the spin-textured SS, which is attributed to continuous filling from a meta-stable population in the CB.  We note that excitation with any above-bandgap photon energy should lead to the same SS filling behavior, since inter- and intra-band scattering processes inevitably lead to relaxation of carriers toward the CB edge.  This phenomenon could find a role in applications requiring ultrafast optical control of a spin-polarized surface conduction channel.  

Finally, we focused on the relaxation of electrons in the CB of $p$-type Bi$_2$Se$_3$.  We found an unexpected decrease of relaxation rates with increasing sample temperature, and showed that this behavior is reproduced by a simple electron-phonon scattering model.  While we have demonstrated this for a nearly-empty band far from $E_F$, we believe these considerations to be equally significant for dynamics near $E_F$.  In fact, it is straightforward to extend this model to this regime by simply removing the assumption that $f\ll1$.  This is not within the scope of the current work, but believe it would be an interesting avenue for future study.  

This analysis serves as an important reminder that the population decay rate extracted by trARPES is not in general equal to the single electron scattering rate.  Rather, the dynamics measured by trARPES is that of a distribution of electrons, which naturally includes a combination of filling and emptying processes from a continuum of states.  In general, the contributions from all co-existing processes must be carefully considered when interpreting transients from trARPES data.

\section{Acknowledgments}
We thank M. Sentef for valuable discussions.  This work is supported by the Department of Energy, Office of Basic Energy Sciences, Division of Materials Science under contract DE-AC02-76SF00515.  J.~A.~S. and S.-L.~Y. acknowledge support by the Stanford Graduate Fellowship.  P.~S.~K. acknowledges support by the Alexander-von-Humboldt foundation through a Feodor-Lynen fellowship and continuous support by M. Wolf.  A.~F.~.K. is supported by the Laboratory Directed Research and Development Program of Lawrence Berkeley National Laboratory under the U.S. Department of Energy contract number DE-AC02-05CH11231. 

\bibliography{Bi2Se3_JESRP}

\begin{thebibliography}{47}%
\makeatletter
\providecommand \@ifxundefined [1]{%
 \@ifx{#1\undefined}
}%
\providecommand \@ifnum [1]{%
 \ifnum #1\expandafter \@firstoftwo
 \else \expandafter \@secondoftwo
 \fi
}%
\providecommand \@ifx [1]{%
 \ifx #1\expandafter \@firstoftwo
 \else \expandafter \@secondoftwo
 \fi
}%
\providecommand \natexlab [1]{#1}%
\providecommand \enquote  [1]{``#1''}%
\providecommand \bibnamefont  [1]{#1}%
\providecommand \bibfnamefont [1]{#1}%
\providecommand \citenamefont [1]{#1}%
\providecommand \href@noop [0]{\@secondoftwo}%
\providecommand \href [0]{\begingroup \@sanitize@url \@href}%
\providecommand \@href[1]{\@@startlink{#1}\@@href}%
\providecommand \@@href[1]{\endgroup#1\@@endlink}%
\providecommand \@sanitize@url [0]{\catcode `\\12\catcode `\$12\catcode
  `\&12\catcode `\#12\catcode `\^12\catcode `\_12\catcode `\%12\relax}%
\providecommand \@@startlink[1]{}%
\providecommand \@@endlink[0]{}%
\providecommand \url  [0]{\begingroup\@sanitize@url \@url }%
\providecommand \@url [1]{\endgroup\@href {#1}{\urlprefix }}%
\providecommand \urlprefix  [0]{URL }%
\providecommand \Eprint [0]{\href }%
\providecommand \doibase [0]{http://dx.doi.org/}%
\providecommand \selectlanguage [0]{\@gobble}%
\providecommand \bibinfo  [0]{\@secondoftwo}%
\providecommand \bibfield  [0]{\@secondoftwo}%
\providecommand \translation [1]{[#1]}%
\providecommand \BibitemOpen [0]{}%
\providecommand \bibitemStop [0]{}%
\providecommand \bibitemNoStop [0]{.\EOS\space}%
\providecommand \EOS [0]{\spacefactor3000\relax}%
\providecommand \BibitemShut  [1]{\csname bibitem#1\endcsname}%
\let\auto@bib@innerbib\@empty
\bibitem [{\citenamefont {Fu}\ \emph {et~al.}(2007)\citenamefont {Fu},
  \citenamefont {Kane},\ and\ \citenamefont {Mele}}]{Fu2007}%
  \BibitemOpen
  \bibfield  {author} {\bibinfo {author} {\bibfnamefont {L.}~\bibnamefont
  {Fu}}, \bibinfo {author} {\bibfnamefont {C.}~\bibnamefont {Kane}}, \ and\
  \bibinfo {author} {\bibfnamefont {E.}~\bibnamefont {Mele}},\ }\href {\doibase
  10.1103/PhysRevLett.98.106803} {\bibfield  {journal} {\bibinfo  {journal}
  {Physical Review Letters}\ }\textbf {\bibinfo {volume} {98}},\ \bibinfo
  {pages} {106803} (\bibinfo {year} {2007})}\BibitemShut {NoStop}%
\bibitem [{\citenamefont {Zhang}\ \emph {et~al.}(2009)\citenamefont {Zhang},
  \citenamefont {Liu}, \citenamefont {Qi}, \citenamefont {Dai}, \citenamefont
  {Fang},\ and\ \citenamefont {Zhang}}]{Zhang2009}%
  \BibitemOpen
  \bibfield  {author} {\bibinfo {author} {\bibfnamefont {H.}~\bibnamefont
  {Zhang}}, \bibinfo {author} {\bibfnamefont {C.-X.}\ \bibnamefont {Liu}},
  \bibinfo {author} {\bibfnamefont {X.-L.}\ \bibnamefont {Qi}}, \bibinfo
  {author} {\bibfnamefont {X.}~\bibnamefont {Dai}}, \bibinfo {author}
  {\bibfnamefont {Z.}~\bibnamefont {Fang}}, \ and\ \bibinfo {author}
  {\bibfnamefont {S.-C.}\ \bibnamefont {Zhang}},\ }\href {\doibase
  10.1038/nphys1270} {\bibfield  {journal} {\bibinfo  {journal} {Nature
  Physics}\ }\textbf {\bibinfo {volume} {5}},\ \bibinfo {pages} {438} (\bibinfo
  {year} {2009})}\BibitemShut {NoStop}%
\bibitem [{\citenamefont {Chen}\ \emph {et~al.}(2009)\citenamefont {Chen},
  \citenamefont {Analytis}, \citenamefont {Chu}, \citenamefont {Liu},
  \citenamefont {Mo}, \citenamefont {Qi}, \citenamefont {Zhang}, \citenamefont
  {Lu}, \citenamefont {Dai}, \citenamefont {Fang}, \citenamefont {Zhang},
  \citenamefont {Fisher}, \citenamefont {Hussain},\ and\ \citenamefont
  {Shen}}]{Chen2009}%
  \BibitemOpen
  \bibfield  {author} {\bibinfo {author} {\bibfnamefont {Y.~L.}\ \bibnamefont
  {Chen}}, \bibinfo {author} {\bibfnamefont {J.~G.}\ \bibnamefont {Analytis}},
  \bibinfo {author} {\bibfnamefont {J.-H.}\ \bibnamefont {Chu}}, \bibinfo
  {author} {\bibfnamefont {Z.~K.}\ \bibnamefont {Liu}}, \bibinfo {author}
  {\bibfnamefont {S.-K.}\ \bibnamefont {Mo}}, \bibinfo {author} {\bibfnamefont
  {X.~L.}\ \bibnamefont {Qi}}, \bibinfo {author} {\bibfnamefont {H.~J.}\
  \bibnamefont {Zhang}}, \bibinfo {author} {\bibfnamefont {D.~H.}\ \bibnamefont
  {Lu}}, \bibinfo {author} {\bibfnamefont {X.}~\bibnamefont {Dai}}, \bibinfo
  {author} {\bibfnamefont {Z.}~\bibnamefont {Fang}}, \bibinfo {author}
  {\bibfnamefont {S.~C.}\ \bibnamefont {Zhang}}, \bibinfo {author}
  {\bibfnamefont {I.~R.}\ \bibnamefont {Fisher}}, \bibinfo {author}
  {\bibfnamefont {Z.}~\bibnamefont {Hussain}}, \ and\ \bibinfo {author}
  {\bibfnamefont {Z.-X.}\ \bibnamefont {Shen}},\ }\href {\doibase
  10.1126/science.1173034} {\bibfield  {journal} {\bibinfo  {journal} {Science
  (New York, N.Y.)}\ }\textbf {\bibinfo {volume} {325}},\ \bibinfo {pages}
  {178} (\bibinfo {year} {2009})}\BibitemShut {NoStop}%
\bibitem [{\citenamefont {Xia}\ \emph {et~al.}(2009)\citenamefont {Xia},
  \citenamefont {Qian}, \citenamefont {Hsieh}, \citenamefont {Wray},
  \citenamefont {Pal}, \citenamefont {Lin}, \citenamefont {Bansil},
  \citenamefont {Grauer}, \citenamefont {Hor}, \citenamefont {Cava},\ and\
  \citenamefont {Hasan}}]{Xia2009}%
  \BibitemOpen
  \bibfield  {author} {\bibinfo {author} {\bibfnamefont {Y.}~\bibnamefont
  {Xia}}, \bibinfo {author} {\bibfnamefont {D.}~\bibnamefont {Qian}}, \bibinfo
  {author} {\bibfnamefont {D.}~\bibnamefont {Hsieh}}, \bibinfo {author}
  {\bibfnamefont {L.}~\bibnamefont {Wray}}, \bibinfo {author} {\bibfnamefont
  {A.}~\bibnamefont {Pal}}, \bibinfo {author} {\bibfnamefont {H.}~\bibnamefont
  {Lin}}, \bibinfo {author} {\bibfnamefont {A.}~\bibnamefont {Bansil}},
  \bibinfo {author} {\bibfnamefont {D.}~\bibnamefont {Grauer}}, \bibinfo
  {author} {\bibfnamefont {Y.~S.}\ \bibnamefont {Hor}}, \bibinfo {author}
  {\bibfnamefont {R.~J.}\ \bibnamefont {Cava}}, \ and\ \bibinfo {author}
  {\bibfnamefont {M.~Z.}\ \bibnamefont {Hasan}},\ }\href {\doibase
  10.1038/nphys1274} {\bibfield  {journal} {\bibinfo  {journal} {Nature
  Physics}\ }\textbf {\bibinfo {volume} {5}},\ \bibinfo {pages} {398} (\bibinfo
  {year} {2009})}\BibitemShut {NoStop}%
\bibitem [{\citenamefont {Qi}\ and\ \citenamefont {Zhang}(2011)}]{Qi2011}%
  \BibitemOpen
  \bibfield  {author} {\bibinfo {author} {\bibfnamefont {X.-L.}\ \bibnamefont
  {Qi}}\ and\ \bibinfo {author} {\bibfnamefont {S.-C.}\ \bibnamefont {Zhang}},\
  }\href {\doibase 10.1103/RevModPhys.83.1057} {\bibfield  {journal} {\bibinfo
  {journal} {Reviews of Modern Physics}\ }\textbf {\bibinfo {volume} {83}},\
  \bibinfo {pages} {1057} (\bibinfo {year} {2011})}\BibitemShut {NoStop}%
\bibitem [{\citenamefont {Hsieh}\ \emph
  {et~al.}(2009{\natexlab{a}})\citenamefont {Hsieh}, \citenamefont {Xia},
  \citenamefont {Qian}, \citenamefont {Wray}, \citenamefont {Dil},
  \citenamefont {Meier}, \citenamefont {Osterwalder}, \citenamefont {Patthey},
  \citenamefont {Checkelsky}, \citenamefont {Ong}, \citenamefont {Fedorov},
  \citenamefont {Lin}, \citenamefont {Bansil}, \citenamefont {Grauer},
  \citenamefont {Hor}, \citenamefont {Cava},\ and\ \citenamefont
  {Hasan}}]{Hsieh2009}%
  \BibitemOpen
  \bibfield  {author} {\bibinfo {author} {\bibfnamefont {D.}~\bibnamefont
  {Hsieh}}, \bibinfo {author} {\bibfnamefont {Y.}~\bibnamefont {Xia}}, \bibinfo
  {author} {\bibfnamefont {D.}~\bibnamefont {Qian}}, \bibinfo {author}
  {\bibfnamefont {L.}~\bibnamefont {Wray}}, \bibinfo {author} {\bibfnamefont
  {J.~H.}\ \bibnamefont {Dil}}, \bibinfo {author} {\bibfnamefont
  {F.}~\bibnamefont {Meier}}, \bibinfo {author} {\bibfnamefont
  {J.}~\bibnamefont {Osterwalder}}, \bibinfo {author} {\bibfnamefont
  {L.}~\bibnamefont {Patthey}}, \bibinfo {author} {\bibfnamefont {J.~G.}\
  \bibnamefont {Checkelsky}}, \bibinfo {author} {\bibfnamefont {N.~P.}\
  \bibnamefont {Ong}}, \bibinfo {author} {\bibfnamefont {a.~V.}\ \bibnamefont
  {Fedorov}}, \bibinfo {author} {\bibfnamefont {H.}~\bibnamefont {Lin}},
  \bibinfo {author} {\bibfnamefont {A.}~\bibnamefont {Bansil}}, \bibinfo
  {author} {\bibfnamefont {D.}~\bibnamefont {Grauer}}, \bibinfo {author}
  {\bibfnamefont {Y.~S.}\ \bibnamefont {Hor}}, \bibinfo {author} {\bibfnamefont
  {R.~J.}\ \bibnamefont {Cava}}, \ and\ \bibinfo {author} {\bibfnamefont
  {M.~Z.}\ \bibnamefont {Hasan}},\ }\href {\doibase 10.1038/nature08234}
  {\bibfield  {journal} {\bibinfo  {journal} {Nature}\ }\textbf {\bibinfo
  {volume} {460}},\ \bibinfo {pages} {1101} (\bibinfo {year}
  {2009}{\natexlab{a}})}\BibitemShut {NoStop}%
\bibitem [{\citenamefont {Hsieh}\ \emph
  {et~al.}(2009{\natexlab{b}})\citenamefont {Hsieh}, \citenamefont {Xia},
  \citenamefont {Wray}, \citenamefont {Qian}, \citenamefont {Pal},
  \citenamefont {Dil}, \citenamefont {Osterwalder}, \citenamefont {Meier},
  \citenamefont {Bihlmayer}, \citenamefont {Kane}, \citenamefont {Hor},
  \citenamefont {Cava},\ and\ \citenamefont {Hasan}}]{Hsieh2009a}%
  \BibitemOpen
  \bibfield  {author} {\bibinfo {author} {\bibfnamefont {D.}~\bibnamefont
  {Hsieh}}, \bibinfo {author} {\bibfnamefont {Y.}~\bibnamefont {Xia}}, \bibinfo
  {author} {\bibfnamefont {L.}~\bibnamefont {Wray}}, \bibinfo {author}
  {\bibfnamefont {D.}~\bibnamefont {Qian}}, \bibinfo {author} {\bibfnamefont
  {A.}~\bibnamefont {Pal}}, \bibinfo {author} {\bibfnamefont {J.~H.}\
  \bibnamefont {Dil}}, \bibinfo {author} {\bibfnamefont {J.}~\bibnamefont
  {Osterwalder}}, \bibinfo {author} {\bibfnamefont {F.}~\bibnamefont {Meier}},
  \bibinfo {author} {\bibfnamefont {G.}~\bibnamefont {Bihlmayer}}, \bibinfo
  {author} {\bibfnamefont {C.~L.}\ \bibnamefont {Kane}}, \bibinfo {author}
  {\bibfnamefont {Y.~S.}\ \bibnamefont {Hor}}, \bibinfo {author} {\bibfnamefont
  {R.~J.}\ \bibnamefont {Cava}}, \ and\ \bibinfo {author} {\bibfnamefont
  {M.~Z.}\ \bibnamefont {Hasan}},\ }\href {\doibase 10.1126/science.1167733}
  {\bibfield  {journal} {\bibinfo  {journal} {Science (New York, N.Y.)}\
  }\textbf {\bibinfo {volume} {323}},\ \bibinfo {pages} {919} (\bibinfo {year}
  {2009}{\natexlab{b}})}\BibitemShut {NoStop}%
\bibitem [{\citenamefont {Roushan}\ \emph {et~al.}(2009)\citenamefont
  {Roushan}, \citenamefont {Seo}, \citenamefont {Parker}, \citenamefont {Hor},
  \citenamefont {Hsieh}, \citenamefont {Qian}, \citenamefont {Richardella},
  \citenamefont {Hasan}, \citenamefont {Cava},\ and\ \citenamefont
  {Yazdani}}]{Roushan2009}%
  \BibitemOpen
  \bibfield  {author} {\bibinfo {author} {\bibfnamefont {P.}~\bibnamefont
  {Roushan}}, \bibinfo {author} {\bibfnamefont {J.}~\bibnamefont {Seo}},
  \bibinfo {author} {\bibfnamefont {C.~V.}\ \bibnamefont {Parker}}, \bibinfo
  {author} {\bibfnamefont {Y.~S.}\ \bibnamefont {Hor}}, \bibinfo {author}
  {\bibfnamefont {D.}~\bibnamefont {Hsieh}}, \bibinfo {author} {\bibfnamefont
  {D.}~\bibnamefont {Qian}}, \bibinfo {author} {\bibfnamefont {A.}~\bibnamefont
  {Richardella}}, \bibinfo {author} {\bibfnamefont {M.~Z.}\ \bibnamefont
  {Hasan}}, \bibinfo {author} {\bibfnamefont {R.~J.}\ \bibnamefont {Cava}}, \
  and\ \bibinfo {author} {\bibfnamefont {A.}~\bibnamefont {Yazdani}},\ }\href
  {\doibase 10.1038/nature08308} {\bibfield  {journal} {\bibinfo  {journal}
  {Nature}\ }\textbf {\bibinfo {volume} {460}},\ \bibinfo {pages} {1106}
  (\bibinfo {year} {2009})}\BibitemShut {NoStop}%
\bibitem [{\citenamefont {Garate}\ and\ \citenamefont
  {Franz}(2010)}]{Garate2010}%
  \BibitemOpen
  \bibfield  {author} {\bibinfo {author} {\bibfnamefont {I.}~\bibnamefont
  {Garate}}\ and\ \bibinfo {author} {\bibfnamefont {M.}~\bibnamefont {Franz}},\
  }\href {\doibase 10.1103/PhysRevLett.104.146802} {\bibfield  {journal}
  {\bibinfo  {journal} {Physical Review Letters}\ }\textbf {\bibinfo {volume}
  {104}},\ \bibinfo {pages} {146802} (\bibinfo {year} {2010})}\BibitemShut
  {NoStop}%
\bibitem [{\citenamefont {Pesin}\ and\ \citenamefont
  {MacDonald}(2012)}]{Pesin2012}%
  \BibitemOpen
  \bibfield  {author} {\bibinfo {author} {\bibfnamefont {D.}~\bibnamefont
  {Pesin}}\ and\ \bibinfo {author} {\bibfnamefont {A.~H.}\ \bibnamefont
  {MacDonald}},\ }\href {\doibase 10.1038/nmat3305} {\bibfield  {journal}
  {\bibinfo  {journal} {Nature materials}\ }\textbf {\bibinfo {volume} {11}},\
  \bibinfo {pages} {409} (\bibinfo {year} {2012})}\BibitemShut {NoStop}%
\bibitem [{\citenamefont {Qi}\ \emph {et~al.}(2010)\citenamefont {Qi},
  \citenamefont {Chen}, \citenamefont {Yu}, \citenamefont {Cadden-Zimansky},
  \citenamefont {Smirnov}, \citenamefont {Tolk}, \citenamefont {Miotkowski},
  \citenamefont {Cao}, \citenamefont {Chen}, \citenamefont {Wu}, \citenamefont
  {Qiao},\ and\ \citenamefont {Jiang}}]{Qi2010}%
  \BibitemOpen
  \bibfield  {author} {\bibinfo {author} {\bibfnamefont {J.}~\bibnamefont
  {Qi}}, \bibinfo {author} {\bibfnamefont {X.}~\bibnamefont {Chen}}, \bibinfo
  {author} {\bibfnamefont {W.}~\bibnamefont {Yu}}, \bibinfo {author}
  {\bibfnamefont {P.}~\bibnamefont {Cadden-Zimansky}}, \bibinfo {author}
  {\bibfnamefont {D.}~\bibnamefont {Smirnov}}, \bibinfo {author} {\bibfnamefont
  {N.~H.}\ \bibnamefont {Tolk}}, \bibinfo {author} {\bibfnamefont
  {I.}~\bibnamefont {Miotkowski}}, \bibinfo {author} {\bibfnamefont
  {H.}~\bibnamefont {Cao}}, \bibinfo {author} {\bibfnamefont {Y.~P.}\
  \bibnamefont {Chen}}, \bibinfo {author} {\bibfnamefont {Y.}~\bibnamefont
  {Wu}}, \bibinfo {author} {\bibfnamefont {S.}~\bibnamefont {Qiao}}, \ and\
  \bibinfo {author} {\bibfnamefont {Z.}~\bibnamefont {Jiang}},\ }\href
  {\doibase 10.1063/1.3513826} {\bibfield  {journal} {\bibinfo  {journal}
  {Applied Physics Letters}\ }\textbf {\bibinfo {volume} {97}},\ \bibinfo
  {pages} {182102} (\bibinfo {year} {2010})}\BibitemShut {NoStop}%
\bibitem [{\citenamefont {Kumar}\ \emph {et~al.}(2011)\citenamefont {Kumar},
  \citenamefont {Ruzicka}, \citenamefont {Butch}, \citenamefont {Syers},
  \citenamefont {Kirshenbaum}, \citenamefont {Paglione},\ and\ \citenamefont
  {Zhao}}]{Kumar2011}%
  \BibitemOpen
  \bibfield  {author} {\bibinfo {author} {\bibfnamefont {N.}~\bibnamefont
  {Kumar}}, \bibinfo {author} {\bibfnamefont {B.~A.}\ \bibnamefont {Ruzicka}},
  \bibinfo {author} {\bibfnamefont {N.~P.}\ \bibnamefont {Butch}}, \bibinfo
  {author} {\bibfnamefont {P.}~\bibnamefont {Syers}}, \bibinfo {author}
  {\bibfnamefont {K.}~\bibnamefont {Kirshenbaum}}, \bibinfo {author}
  {\bibfnamefont {J.}~\bibnamefont {Paglione}}, \ and\ \bibinfo {author}
  {\bibfnamefont {H.}~\bibnamefont {Zhao}},\ }\href {\doibase
  10.1103/PhysRevB.83.235306} {\bibfield  {journal} {\bibinfo  {journal}
  {Physical Review B}\ }\textbf {\bibinfo {volume} {83}},\ \bibinfo {pages}
  {235306} (\bibinfo {year} {2011})}\BibitemShut {NoStop}%
\bibitem [{\citenamefont {Glinka}\ \emph {et~al.}(2013)\citenamefont {Glinka},
  \citenamefont {Babakiray}, \citenamefont {Johnson}, \citenamefont {Bristow},
  \citenamefont {Holcomb},\ and\ \citenamefont {Lederman}}]{Glinka2013}%
  \BibitemOpen
  \bibfield  {author} {\bibinfo {author} {\bibfnamefont {Y.~D.}\ \bibnamefont
  {Glinka}}, \bibinfo {author} {\bibfnamefont {S.}~\bibnamefont {Babakiray}},
  \bibinfo {author} {\bibfnamefont {T.~A.}\ \bibnamefont {Johnson}}, \bibinfo
  {author} {\bibfnamefont {A.~D.}\ \bibnamefont {Bristow}}, \bibinfo {author}
  {\bibfnamefont {M.~B.}\ \bibnamefont {Holcomb}}, \ and\ \bibinfo {author}
  {\bibfnamefont {D.}~\bibnamefont {Lederman}},\ }\href {\doibase
  10.1063/1.4824821} {\bibfield  {journal} {\bibinfo  {journal} {Applied
  Physics Letters}\ }\textbf {\bibinfo {volume} {103}},\ \bibinfo {pages}
  {151903} (\bibinfo {year} {2013})}\BibitemShut {NoStop}%
\bibitem [{\citenamefont {Hsieh}\ \emph {et~al.}(2011)\citenamefont {Hsieh},
  \citenamefont {Mahmood}, \citenamefont {McIver}, \citenamefont {Gardner},
  \citenamefont {Lee},\ and\ \citenamefont {Gedik}}]{Hsieh2011a}%
  \BibitemOpen
  \bibfield  {author} {\bibinfo {author} {\bibfnamefont {D.}~\bibnamefont
  {Hsieh}}, \bibinfo {author} {\bibfnamefont {F.}~\bibnamefont {Mahmood}},
  \bibinfo {author} {\bibfnamefont {J.~W.}\ \bibnamefont {McIver}}, \bibinfo
  {author} {\bibfnamefont {D.~R.}\ \bibnamefont {Gardner}}, \bibinfo {author}
  {\bibfnamefont {Y.~S.}\ \bibnamefont {Lee}}, \ and\ \bibinfo {author}
  {\bibfnamefont {N.}~\bibnamefont {Gedik}},\ }\href {\doibase
  10.1103/PhysRevLett.107.077401} {\bibfield  {journal} {\bibinfo  {journal}
  {Physical Review Letters}\ }\textbf {\bibinfo {volume} {107}},\ \bibinfo
  {pages} {077401} (\bibinfo {year} {2011})}\BibitemShut {NoStop}%
\bibitem [{\citenamefont {Sobota}\ \emph {et~al.}(2012)\citenamefont {Sobota},
  \citenamefont {Yang}, \citenamefont {Analytis}, \citenamefont {Chen},
  \citenamefont {Fisher}, \citenamefont {Kirchmann},\ and\ \citenamefont
  {Shen}}]{Sobota2012}%
  \BibitemOpen
  \bibfield  {author} {\bibinfo {author} {\bibfnamefont {J.~A.}\ \bibnamefont
  {Sobota}}, \bibinfo {author} {\bibfnamefont {S.}~\bibnamefont {Yang}},
  \bibinfo {author} {\bibfnamefont {J.~G.}\ \bibnamefont {Analytis}}, \bibinfo
  {author} {\bibfnamefont {Y.~L.}\ \bibnamefont {Chen}}, \bibinfo {author}
  {\bibfnamefont {I.~R.}\ \bibnamefont {Fisher}}, \bibinfo {author}
  {\bibfnamefont {P.~S.}\ \bibnamefont {Kirchmann}}, \ and\ \bibinfo {author}
  {\bibfnamefont {Z.-X.}\ \bibnamefont {Shen}},\ }\href {\doibase
  10.1103/PhysRevLett.108.117403} {\bibfield  {journal} {\bibinfo  {journal}
  {Physical Review Letters}\ }\textbf {\bibinfo {volume} {108}},\ \bibinfo
  {pages} {117403} (\bibinfo {year} {2012})}\BibitemShut {NoStop}%
\bibitem [{\citenamefont {Hajlaoui}\ \emph {et~al.}(2012)\citenamefont
  {Hajlaoui}, \citenamefont {Papalazarou}, \citenamefont {Mauchain},
  \citenamefont {Lantz}, \citenamefont {Moisan}, \citenamefont {Boschetto},
  \citenamefont {Jiang}, \citenamefont {Miotkowski}, \citenamefont {Chen},
  \citenamefont {Taleb-Ibrahimi}, \citenamefont {Perfetti},\ and\ \citenamefont
  {Marsi}}]{Hajlaoui2012}%
  \BibitemOpen
  \bibfield  {author} {\bibinfo {author} {\bibfnamefont {M.}~\bibnamefont
  {Hajlaoui}}, \bibinfo {author} {\bibfnamefont {E.}~\bibnamefont
  {Papalazarou}}, \bibinfo {author} {\bibfnamefont {J.}~\bibnamefont
  {Mauchain}}, \bibinfo {author} {\bibfnamefont {G.}~\bibnamefont {Lantz}},
  \bibinfo {author} {\bibfnamefont {N.}~\bibnamefont {Moisan}}, \bibinfo
  {author} {\bibfnamefont {D.}~\bibnamefont {Boschetto}}, \bibinfo {author}
  {\bibfnamefont {Z.}~\bibnamefont {Jiang}}, \bibinfo {author} {\bibfnamefont
  {I.}~\bibnamefont {Miotkowski}}, \bibinfo {author} {\bibfnamefont {Y.~P.}\
  \bibnamefont {Chen}}, \bibinfo {author} {\bibfnamefont {A.}~\bibnamefont
  {Taleb-Ibrahimi}}, \bibinfo {author} {\bibfnamefont {L.}~\bibnamefont
  {Perfetti}}, \ and\ \bibinfo {author} {\bibfnamefont {M.}~\bibnamefont
  {Marsi}},\ }\href {\doibase 10.1021/nl301035x} {\bibfield  {journal}
  {\bibinfo  {journal} {Nano letters}\ }\textbf {\bibinfo {volume} {12}},\
  \bibinfo {pages} {3532} (\bibinfo {year} {2012})}\BibitemShut {NoStop}%
\bibitem [{\citenamefont {Wang}\ \emph {et~al.}(2012)\citenamefont {Wang},
  \citenamefont {Hsieh}, \citenamefont {Sie}, \citenamefont {Steinberg},
  \citenamefont {Gardner}, \citenamefont {Lee}, \citenamefont
  {Jarillo-Herrero},\ and\ \citenamefont {Gedik}}]{Wang2012a}%
  \BibitemOpen
  \bibfield  {author} {\bibinfo {author} {\bibfnamefont {Y.~H.}\ \bibnamefont
  {Wang}}, \bibinfo {author} {\bibfnamefont {D.}~\bibnamefont {Hsieh}},
  \bibinfo {author} {\bibfnamefont {E.~J.}\ \bibnamefont {Sie}}, \bibinfo
  {author} {\bibfnamefont {H.}~\bibnamefont {Steinberg}}, \bibinfo {author}
  {\bibfnamefont {D.~R.}\ \bibnamefont {Gardner}}, \bibinfo {author}
  {\bibfnamefont {Y.~S.}\ \bibnamefont {Lee}}, \bibinfo {author} {\bibfnamefont
  {P.}~\bibnamefont {Jarillo-Herrero}}, \ and\ \bibinfo {author} {\bibfnamefont
  {N.}~\bibnamefont {Gedik}},\ }\href {\doibase 10.1103/PhysRevLett.109.127401}
  {\bibfield  {journal} {\bibinfo  {journal} {Physical Review Letters}\
  }\textbf {\bibinfo {volume} {109}},\ \bibinfo {pages} {127401} (\bibinfo
  {year} {2012})}\BibitemShut {NoStop}%
\bibitem [{\citenamefont {Hajlaoui}\ \emph {et~al.}(2013)\citenamefont
  {Hajlaoui}, \citenamefont {Papalazarou}, \citenamefont {Mauchain},
  \citenamefont {Jiang}, \citenamefont {Miotkowski}, \citenamefont {Chen},
  \citenamefont {Taleb-Ibrahimi}, \citenamefont {Perfetti},\ and\ \citenamefont
  {Marsi}}]{Hajlaoui2013}%
  \BibitemOpen
  \bibfield  {author} {\bibinfo {author} {\bibfnamefont {M.}~\bibnamefont
  {Hajlaoui}}, \bibinfo {author} {\bibfnamefont {E.}~\bibnamefont
  {Papalazarou}}, \bibinfo {author} {\bibfnamefont {J.}~\bibnamefont
  {Mauchain}}, \bibinfo {author} {\bibfnamefont {Z.}~\bibnamefont {Jiang}},
  \bibinfo {author} {\bibfnamefont {I.}~\bibnamefont {Miotkowski}}, \bibinfo
  {author} {\bibfnamefont {Y.~P.}\ \bibnamefont {Chen}}, \bibinfo {author}
  {\bibfnamefont {A.}~\bibnamefont {Taleb-Ibrahimi}}, \bibinfo {author}
  {\bibfnamefont {L.}~\bibnamefont {Perfetti}}, \ and\ \bibinfo {author}
  {\bibfnamefont {M.}~\bibnamefont {Marsi}},\ }\href {\doibase
  10.1140/epjst/e2013-01921-1} {\bibfield  {journal} {\bibinfo  {journal} {The
  European Physical Journal Special Topics}\ }\textbf {\bibinfo {volume}
  {222}},\ \bibinfo {pages} {1271} (\bibinfo {year} {2013})}\BibitemShut
  {NoStop}%
\bibitem [{\citenamefont {Crepaldi}\ \emph {et~al.}(2012)\citenamefont
  {Crepaldi}, \citenamefont {Ressel}, \citenamefont {Cilento}, \citenamefont
  {Zacchigna}, \citenamefont {Grazioli}, \citenamefont {Berger}, \citenamefont
  {Bugnon}, \citenamefont {Kern}, \citenamefont {Grioni},\ and\ \citenamefont
  {Parmigiani}}]{Crepaldi2012}%
  \BibitemOpen
  \bibfield  {author} {\bibinfo {author} {\bibfnamefont {A.}~\bibnamefont
  {Crepaldi}}, \bibinfo {author} {\bibfnamefont {B.}~\bibnamefont {Ressel}},
  \bibinfo {author} {\bibfnamefont {F.}~\bibnamefont {Cilento}}, \bibinfo
  {author} {\bibfnamefont {M.}~\bibnamefont {Zacchigna}}, \bibinfo {author}
  {\bibfnamefont {C.}~\bibnamefont {Grazioli}}, \bibinfo {author}
  {\bibfnamefont {H.}~\bibnamefont {Berger}}, \bibinfo {author} {\bibfnamefont
  {P.}~\bibnamefont {Bugnon}}, \bibinfo {author} {\bibfnamefont
  {K.}~\bibnamefont {Kern}}, \bibinfo {author} {\bibfnamefont {M.}~\bibnamefont
  {Grioni}}, \ and\ \bibinfo {author} {\bibfnamefont {F.}~\bibnamefont
  {Parmigiani}},\ }\href {\doibase 10.1103/PhysRevB.86.205133} {\bibfield
  {journal} {\bibinfo  {journal} {Physical Review B}\ }\textbf {\bibinfo
  {volume} {86}},\ \bibinfo {pages} {205133} (\bibinfo {year}
  {2012})}\BibitemShut {NoStop}%
\bibitem [{\citenamefont {Crepaldi}\ \emph {et~al.}(2013)\citenamefont
  {Crepaldi}, \citenamefont {Cilento}, \citenamefont {Ressel}, \citenamefont
  {Cacho}, \citenamefont {Johannsen}, \citenamefont {Zacchigna}, \citenamefont
  {Berger}, \citenamefont {Bugnon}, \citenamefont {Grazioli}, \citenamefont
  {Turcu}, \citenamefont {Springate}, \citenamefont {Kern}, \citenamefont
  {Grioni},\ and\ \citenamefont {Parmigiani}}]{Crepaldi2013}%
  \BibitemOpen
  \bibfield  {author} {\bibinfo {author} {\bibfnamefont {A.}~\bibnamefont
  {Crepaldi}}, \bibinfo {author} {\bibfnamefont {F.}~\bibnamefont {Cilento}},
  \bibinfo {author} {\bibfnamefont {B.}~\bibnamefont {Ressel}}, \bibinfo
  {author} {\bibfnamefont {C.}~\bibnamefont {Cacho}}, \bibinfo {author}
  {\bibfnamefont {J.~C.}\ \bibnamefont {Johannsen}}, \bibinfo {author}
  {\bibfnamefont {M.}~\bibnamefont {Zacchigna}}, \bibinfo {author}
  {\bibfnamefont {H.}~\bibnamefont {Berger}}, \bibinfo {author} {\bibfnamefont
  {P.}~\bibnamefont {Bugnon}}, \bibinfo {author} {\bibfnamefont
  {C.}~\bibnamefont {Grazioli}}, \bibinfo {author} {\bibfnamefont {I.~C.~E.}\
  \bibnamefont {Turcu}}, \bibinfo {author} {\bibfnamefont {E.}~\bibnamefont
  {Springate}}, \bibinfo {author} {\bibfnamefont {K.}~\bibnamefont {Kern}},
  \bibinfo {author} {\bibfnamefont {M.}~\bibnamefont {Grioni}}, \ and\ \bibinfo
  {author} {\bibfnamefont {F.}~\bibnamefont {Parmigiani}},\ }\href {\doibase
  10.1103/PhysRevB.88.121404} {\bibfield  {journal} {\bibinfo  {journal}
  {Physical Review B}\ }\textbf {\bibinfo {volume} {88}},\ \bibinfo {pages}
  {121404} (\bibinfo {year} {2013})}\BibitemShut {NoStop}%
\bibitem [{\citenamefont {Niesner}\ \emph {et~al.}(2012)\citenamefont
  {Niesner}, \citenamefont {Fauster}, \citenamefont {Eremeev}, \citenamefont
  {Menshchikova}, \citenamefont {Koroteev}, \citenamefont {Protogenov},
  \citenamefont {Chulkov}, \citenamefont {Tereshchenko}, \citenamefont {Kokh},
  \citenamefont {Alekperov}, \citenamefont {Nadjafov},\ and\ \citenamefont
  {Mamedov}}]{Niesner2012}%
  \BibitemOpen
  \bibfield  {author} {\bibinfo {author} {\bibfnamefont {D.}~\bibnamefont
  {Niesner}}, \bibinfo {author} {\bibfnamefont {T.}~\bibnamefont {Fauster}},
  \bibinfo {author} {\bibfnamefont {S.~V.}\ \bibnamefont {Eremeev}}, \bibinfo
  {author} {\bibfnamefont {T.~V.}\ \bibnamefont {Menshchikova}}, \bibinfo
  {author} {\bibfnamefont {Y.~M.}\ \bibnamefont {Koroteev}}, \bibinfo {author}
  {\bibfnamefont {A.~P.}\ \bibnamefont {Protogenov}}, \bibinfo {author}
  {\bibfnamefont {E.~V.}\ \bibnamefont {Chulkov}}, \bibinfo {author}
  {\bibfnamefont {O.~E.}\ \bibnamefont {Tereshchenko}}, \bibinfo {author}
  {\bibfnamefont {K.~A.}\ \bibnamefont {Kokh}}, \bibinfo {author}
  {\bibfnamefont {O.}~\bibnamefont {Alekperov}}, \bibinfo {author}
  {\bibfnamefont {A.}~\bibnamefont {Nadjafov}}, \ and\ \bibinfo {author}
  {\bibfnamefont {N.}~\bibnamefont {Mamedov}},\ }\href {\doibase
  10.1103/PhysRevB.86.205403} {\bibfield  {journal} {\bibinfo  {journal}
  {Physical Review B}\ }\textbf {\bibinfo {volume} {86}},\ \bibinfo {pages}
  {205403} (\bibinfo {year} {2012})}\BibitemShut {NoStop}%
\bibitem [{\citenamefont {Sobota}\ \emph {et~al.}(2013)\citenamefont {Sobota},
  \citenamefont {Yang}, \citenamefont {Kemper}, \citenamefont {Lee},
  \citenamefont {Schmitt}, \citenamefont {Li}, \citenamefont {Moore},
  \citenamefont {Analytis}, \citenamefont {Fisher}, \citenamefont {Kirchmann},
  \citenamefont {Devereaux},\ and\ \citenamefont {Shen}}]{Sobota2013}%
  \BibitemOpen
  \bibfield  {author} {\bibinfo {author} {\bibfnamefont {J.~A.}\ \bibnamefont
  {Sobota}}, \bibinfo {author} {\bibfnamefont {S.-L.}\ \bibnamefont {Yang}},
  \bibinfo {author} {\bibfnamefont {A.~F.}\ \bibnamefont {Kemper}}, \bibinfo
  {author} {\bibfnamefont {J.~J.}\ \bibnamefont {Lee}}, \bibinfo {author}
  {\bibfnamefont {F.~T.}\ \bibnamefont {Schmitt}}, \bibinfo {author}
  {\bibfnamefont {W.}~\bibnamefont {Li}}, \bibinfo {author} {\bibfnamefont
  {R.~G.}\ \bibnamefont {Moore}}, \bibinfo {author} {\bibfnamefont {J.~G.}\
  \bibnamefont {Analytis}}, \bibinfo {author} {\bibfnamefont {I.~R.}\
  \bibnamefont {Fisher}}, \bibinfo {author} {\bibfnamefont {P.~S.}\
  \bibnamefont {Kirchmann}}, \bibinfo {author} {\bibfnamefont {T.~P.}\
  \bibnamefont {Devereaux}}, \ and\ \bibinfo {author} {\bibfnamefont {Z.-X.}\
  \bibnamefont {Shen}},\ }\href {\doibase 10.1103/PhysRevLett.111.136802}
  {\bibfield  {journal} {\bibinfo  {journal} {Physical Review Letters}\
  }\textbf {\bibinfo {volume} {111}},\ \bibinfo {pages} {136802} (\bibinfo
  {year} {2013})}\BibitemShut {NoStop}%
\bibitem [{\citenamefont {Wang}\ \emph {et~al.}(2013)\citenamefont {Wang},
  \citenamefont {Steinberg}, \citenamefont {Jarillo-Herrero},\ and\
  \citenamefont {Gedik}}]{Wang2013}%
  \BibitemOpen
  \bibfield  {author} {\bibinfo {author} {\bibfnamefont {Y.~H.}\ \bibnamefont
  {Wang}}, \bibinfo {author} {\bibfnamefont {H.}~\bibnamefont {Steinberg}},
  \bibinfo {author} {\bibfnamefont {P.}~\bibnamefont {Jarillo-Herrero}}, \ and\
  \bibinfo {author} {\bibfnamefont {N.}~\bibnamefont {Gedik}},\ }\href
  {\doibase 10.1126/science.1239834} {\bibfield  {journal} {\bibinfo  {journal}
  {Science (New York, N.Y.)}\ }\textbf {\bibinfo {volume} {342}},\ \bibinfo
  {pages} {453} (\bibinfo {year} {2013})}\BibitemShut {NoStop}%
\bibitem [{\citenamefont {Haight}(1995)}]{Haight1995}%
  \BibitemOpen
  \bibfield  {author} {\bibinfo {author} {\bibfnamefont {R.}~\bibnamefont
  {Haight}},\ }\href {\doibase 10.1016/0167-5729(95)00002-X} {\bibfield
  {journal} {\bibinfo  {journal} {Surface Science Reports}\ }\textbf {\bibinfo
  {volume} {21}},\ \bibinfo {pages} {275} (\bibinfo {year} {1995})}\BibitemShut
  {NoStop}%
\bibitem [{\citenamefont {Petek}\ and\ \citenamefont
  {Ogawa}(1997)}]{Petek1998}%
  \BibitemOpen
  \bibfield  {author} {\bibinfo {author} {\bibfnamefont {H.}~\bibnamefont
  {Petek}}\ and\ \bibinfo {author} {\bibfnamefont {S.}~\bibnamefont {Ogawa}},\
  }\href {\doibase 10.1016/S0079-6816(98)00002-1} {\bibfield  {journal}
  {\bibinfo  {journal} {Progress in Surface Science}\ }\textbf {\bibinfo
  {volume} {56}},\ \bibinfo {pages} {239} (\bibinfo {year} {1997})}\BibitemShut
  {NoStop}%
\bibitem [{\citenamefont {Weinelt}(2002)}]{Weinelt2002}%
  \BibitemOpen
  \bibfield  {author} {\bibinfo {author} {\bibfnamefont {M.}~\bibnamefont
  {Weinelt}},\ }\href {\doibase 10.1088/0953-8984/14/43/202} {\bibfield
  {journal} {\bibinfo  {journal} {Journal of Physics: Condensed Matter}\
  }\textbf {\bibinfo {volume} {14}},\ \bibinfo {pages} {R1099} (\bibinfo {year}
  {2002})}\BibitemShut {NoStop}%
\bibitem [{\citenamefont {H\"{u}fner}(1995)}]{Hufner1995}%
  \BibitemOpen
  \bibfield  {author} {\bibinfo {author} {\bibfnamefont {S.}~\bibnamefont
  {H\"{u}fner}},\ }\href@noop {} {\emph {\bibinfo {title} {{Photoelectron
  Spectroscopy}}}}\ (\bibinfo  {publisher} {Springer},\ \bibinfo {address}
  {Berlin},\ \bibinfo {year} {1995})\BibitemShut {NoStop}%
\bibitem [{\citenamefont {Perdew}\ \emph {et~al.}(1996)\citenamefont {Perdew},
  \citenamefont {Burke},\ and\ \citenamefont {Ernzerhof}}]{Perdew1996}%
  \BibitemOpen
  \bibfield  {author} {\bibinfo {author} {\bibfnamefont {J.~P.}\ \bibnamefont
  {Perdew}}, \bibinfo {author} {\bibfnamefont {K.}~\bibnamefont {Burke}}, \
  and\ \bibinfo {author} {\bibfnamefont {M.}~\bibnamefont {Ernzerhof}},\ }\href
  {\doibase 10.1103/PhysRevLett.77.3865} {\bibfield  {journal} {\bibinfo
  {journal} {Physical Review Letters}\ }\textbf {\bibinfo {volume} {77}},\
  \bibinfo {pages} {3865} (\bibinfo {year} {1996})}\BibitemShut {NoStop}%
\bibitem [{\citenamefont {Blaha}\ \emph {et~al.}(2001)\citenamefont {Blaha},
  \citenamefont {Schwarz}, \citenamefont {Madsen}, \citenamefont {Kvasnicka},\
  and\ \citenamefont {Luitz}}]{Blaha2001}%
  \BibitemOpen
  \bibfield  {author} {\bibinfo {author} {\bibfnamefont {P.}~\bibnamefont
  {Blaha}}, \bibinfo {author} {\bibfnamefont {K.}~\bibnamefont {Schwarz}},
  \bibinfo {author} {\bibfnamefont {G.}~\bibnamefont {Madsen}}, \bibinfo
  {author} {\bibfnamefont {D.}~\bibnamefont {Kvasnicka}}, \ and\ \bibinfo
  {author} {\bibfnamefont {J.}~\bibnamefont {Luitz}},\ }\href@noop {} {\emph
  {\bibinfo {title} {{WIEN2k, An Augmented Plane Wave Plus Local Orbitals
  Program for Calculating Crystal Properties}}}},\ edited by\ \bibinfo {editor}
  {\bibfnamefont {K.}~\bibnamefont {Schwarz}}\ (\bibinfo  {publisher} {Techn.
  Universitat Wein, Austria},\ \bibinfo {year} {2001})\BibitemShut {NoStop}%
\bibitem [{\citenamefont {Nakajima}(1963)}]{Nakajima1963}%
  \BibitemOpen
  \bibfield  {author} {\bibinfo {author} {\bibfnamefont {S.}~\bibnamefont
  {Nakajima}},\ }\href {\doibase 10.1016/0022-3697(63)90207-5} {\bibfield
  {journal} {\bibinfo  {journal} {Journal of Physics and Chemistry of Solids}\
  }\textbf {\bibinfo {volume} {24}},\ \bibinfo {pages} {479} (\bibinfo {year}
  {1963})}\BibitemShut {NoStop}%
\bibitem [{\citenamefont {Giuliani}\ and\ \citenamefont
  {Vignale}(2005)}]{Giu05}%
  \BibitemOpen
  \bibfield  {author} {\bibinfo {author} {\bibfnamefont {G.~F.}\ \bibnamefont
  {Giuliani}}\ and\ \bibinfo {author} {\bibfnamefont {G.}~\bibnamefont
  {Vignale}},\ }\href {\doibase 10.2277} {\emph {\bibinfo {title} {{Quantum
  Theory of the Electron Liquid}}}}\ (\bibinfo  {publisher} {Cambridge
  University Press},\ \bibinfo {address} {Cambridge, U.K.},\ \bibinfo {year}
  {2005})\BibitemShut {NoStop}%
\bibitem [{\citenamefont {Kuroda}\ \emph {et~al.}(2010)\citenamefont {Kuroda},
  \citenamefont {Arita}, \citenamefont {Miyamoto}, \citenamefont {Ye},
  \citenamefont {Jiang}, \citenamefont {Kimura}, \citenamefont {Krasovskii},
  \citenamefont {Chulkov}, \citenamefont {Iwasawa}, \citenamefont {Okuda},
  \citenamefont {Shimada}, \citenamefont {Ueda}, \citenamefont {Namatame},\
  and\ \citenamefont {Taniguchi}}]{Kuroda2010}%
  \BibitemOpen
  \bibfield  {author} {\bibinfo {author} {\bibfnamefont {K.}~\bibnamefont
  {Kuroda}}, \bibinfo {author} {\bibfnamefont {M.}~\bibnamefont {Arita}},
  \bibinfo {author} {\bibfnamefont {K.}~\bibnamefont {Miyamoto}}, \bibinfo
  {author} {\bibfnamefont {M.}~\bibnamefont {Ye}}, \bibinfo {author}
  {\bibfnamefont {J.}~\bibnamefont {Jiang}}, \bibinfo {author} {\bibfnamefont
  {A.}~\bibnamefont {Kimura}}, \bibinfo {author} {\bibfnamefont {E.~E.}\
  \bibnamefont {Krasovskii}}, \bibinfo {author} {\bibfnamefont {E.~V.}\
  \bibnamefont {Chulkov}}, \bibinfo {author} {\bibfnamefont {H.}~\bibnamefont
  {Iwasawa}}, \bibinfo {author} {\bibfnamefont {T.}~\bibnamefont {Okuda}},
  \bibinfo {author} {\bibfnamefont {K.}~\bibnamefont {Shimada}}, \bibinfo
  {author} {\bibfnamefont {Y.}~\bibnamefont {Ueda}}, \bibinfo {author}
  {\bibfnamefont {H.}~\bibnamefont {Namatame}}, \ and\ \bibinfo {author}
  {\bibfnamefont {M.}~\bibnamefont {Taniguchi}},\ }\href {\doibase
  10.1103/PhysRevLett.105.076802} {\bibfield  {journal} {\bibinfo  {journal}
  {Physical Review Letters}\ }\textbf {\bibinfo {volume} {105}},\ \bibinfo
  {pages} {076802} (\bibinfo {year} {2010})}\BibitemShut {NoStop}%
\bibitem [{\citenamefont {Eremeev}\ \emph {et~al.}(2013)\citenamefont
  {Eremeev}, \citenamefont {Silkin}, \citenamefont {Menshchikova},
  \citenamefont {Protogenov},\ and\ \citenamefont {Chulkov}}]{Eremeev2013}%
  \BibitemOpen
  \bibfield  {author} {\bibinfo {author} {\bibfnamefont {S.~V.}\ \bibnamefont
  {Eremeev}}, \bibinfo {author} {\bibfnamefont {I.~V.}\ \bibnamefont {Silkin}},
  \bibinfo {author} {\bibfnamefont {T.~V.}\ \bibnamefont {Menshchikova}},
  \bibinfo {author} {\bibfnamefont {A.~P.}\ \bibnamefont {Protogenov}}, \ and\
  \bibinfo {author} {\bibfnamefont {E.~V.}\ \bibnamefont {Chulkov}},\ }\href
  {\doibase 10.1134/S0021364012240034} {\bibfield  {journal} {\bibinfo
  {journal} {JETP Letters}\ }\textbf {\bibinfo {volume} {96}},\ \bibinfo
  {pages} {780} (\bibinfo {year} {2013})}\BibitemShut {NoStop}%
\bibitem [{\citenamefont {Sze}\ and\ \citenamefont {Ng}(2006)}]{Sze2006}%
  \BibitemOpen
  \bibfield  {author} {\bibinfo {author} {\bibfnamefont {S.}~\bibnamefont
  {Sze}}\ and\ \bibinfo {author} {\bibfnamefont {K.~K.}\ \bibnamefont {Ng}},\
  }\href {\doibase 10.1002/0470068329} {\emph {\bibinfo {title}
  {Engineering}}}\ (\bibinfo  {publisher} {John Wiley \& Sons, Inc.},\ \bibinfo
  {address} {Hoboken, NJ, USA},\ \bibinfo {year} {2006})\BibitemShut {NoStop}%
\bibitem [{\citenamefont {Richter}\ and\ \citenamefont
  {Becker}(1977)}]{Richter1977}%
  \BibitemOpen
  \bibfield  {author} {\bibinfo {author} {\bibfnamefont {W.}~\bibnamefont
  {Richter}}\ and\ \bibinfo {author} {\bibfnamefont {C.~R.}\ \bibnamefont
  {Becker}},\ }\href {\doibase 10.1002/pssb.2220840226} {\bibfield  {journal}
  {\bibinfo  {journal} {Physica Status Solidi (b)}\ }\textbf {\bibinfo {volume}
  {84}},\ \bibinfo {pages} {619} (\bibinfo {year} {1977})}\BibitemShut
  {NoStop}%
\bibitem [{\citenamefont {Chen}\ \emph {et~al.}(2010)\citenamefont {Chen},
  \citenamefont {Chu}, \citenamefont {Analytis}, \citenamefont {Liu},
  \citenamefont {Igarashi}, \citenamefont {Kuo}, \citenamefont {Qi},
  \citenamefont {Mo}, \citenamefont {Moore}, \citenamefont {Lu}, \citenamefont
  {Hashimoto}, \citenamefont {Sasagawa}, \citenamefont {Zhang}, \citenamefont
  {Fisher}, \citenamefont {Hussain},\ and\ \citenamefont {Shen}}]{Chen2010}%
  \BibitemOpen
  \bibfield  {author} {\bibinfo {author} {\bibfnamefont {Y.~L.}\ \bibnamefont
  {Chen}}, \bibinfo {author} {\bibfnamefont {J.-H.}\ \bibnamefont {Chu}},
  \bibinfo {author} {\bibfnamefont {J.~G.}\ \bibnamefont {Analytis}}, \bibinfo
  {author} {\bibfnamefont {Z.~K.}\ \bibnamefont {Liu}}, \bibinfo {author}
  {\bibfnamefont {K.}~\bibnamefont {Igarashi}}, \bibinfo {author}
  {\bibfnamefont {H.-H.}\ \bibnamefont {Kuo}}, \bibinfo {author} {\bibfnamefont
  {X.~L.}\ \bibnamefont {Qi}}, \bibinfo {author} {\bibfnamefont {S.~K.}\
  \bibnamefont {Mo}}, \bibinfo {author} {\bibfnamefont {R.~G.}\ \bibnamefont
  {Moore}}, \bibinfo {author} {\bibfnamefont {D.~H.}\ \bibnamefont {Lu}},
  \bibinfo {author} {\bibfnamefont {M.}~\bibnamefont {Hashimoto}}, \bibinfo
  {author} {\bibfnamefont {T.}~\bibnamefont {Sasagawa}}, \bibinfo {author}
  {\bibfnamefont {S.~C.}\ \bibnamefont {Zhang}}, \bibinfo {author}
  {\bibfnamefont {I.~R.}\ \bibnamefont {Fisher}}, \bibinfo {author}
  {\bibfnamefont {Z.}~\bibnamefont {Hussain}}, \ and\ \bibinfo {author}
  {\bibfnamefont {Z.~X.}\ \bibnamefont {Shen}},\ }\href {\doibase
  10.1126/science.1189924} {\bibfield  {journal} {\bibinfo  {journal} {Science
  (New York, N.Y.)}\ }\textbf {\bibinfo {volume} {329}},\ \bibinfo {pages}
  {659} (\bibinfo {year} {2010})}\BibitemShut {NoStop}%
\bibitem [{\citenamefont {Weinelt}\ \emph {et~al.}(2004)\citenamefont
  {Weinelt}, \citenamefont {Kutschera}, \citenamefont {Fauster},\ and\
  \citenamefont {Rohlfing}}]{Weinelt2004}%
  \BibitemOpen
  \bibfield  {author} {\bibinfo {author} {\bibfnamefont {M.}~\bibnamefont
  {Weinelt}}, \bibinfo {author} {\bibfnamefont {M.}~\bibnamefont {Kutschera}},
  \bibinfo {author} {\bibfnamefont {T.}~\bibnamefont {Fauster}}, \ and\
  \bibinfo {author} {\bibfnamefont {M.}~\bibnamefont {Rohlfing}},\ }\href
  {\doibase 10.1103/PhysRevLett.92.126801} {\bibfield  {journal} {\bibinfo
  {journal} {Physical Review Letters}\ }\textbf {\bibinfo {volume} {92}},\
  \bibinfo {pages} {90} (\bibinfo {year} {2004})}\BibitemShut {NoStop}%
\bibitem [{\citenamefont {Tanaka}\ \emph {et~al.}(2009)\citenamefont {Tanaka},
  \citenamefont {Ichibayashi},\ and\ \citenamefont {Tanimura}}]{Tanaka2009}%
  \BibitemOpen
  \bibfield  {author} {\bibinfo {author} {\bibfnamefont {S.}~\bibnamefont
  {Tanaka}}, \bibinfo {author} {\bibfnamefont {T.}~\bibnamefont {Ichibayashi}},
  \ and\ \bibinfo {author} {\bibfnamefont {K.}~\bibnamefont {Tanimura}},\
  }\href {\doibase 10.1103/PhysRevB.79.155313} {\bibfield  {journal} {\bibinfo
  {journal} {Physical Review B}\ }\textbf {\bibinfo {volume} {79}},\ \bibinfo
  {pages} {155313} (\bibinfo {year} {2009})}\BibitemShut {NoStop}%
\bibitem [{\citenamefont {Schroder}(2005)}]{Schroder2005}%
  \BibitemOpen
  \bibfield  {author} {\bibinfo {author} {\bibfnamefont {D.~K.}\ \bibnamefont
  {Schroder}},\ }\href {\doibase 10.1002/0471749095} {\emph {\bibinfo {title}
  {{Semiconductor Material and Device Characterization}}}}\ (\bibinfo
  {publisher} {John Wiley \& Sons, Inc.},\ \bibinfo {address} {Hoboken, NJ,
  USA},\ \bibinfo {year} {2005})\BibitemShut {NoStop}%
\bibitem [{\citenamefont {Grimvall}(1981)}]{Grimvall1981}%
  \BibitemOpen
  \bibfield  {author} {\bibinfo {author} {\bibfnamefont {G.}~\bibnamefont
  {Grimvall}},\ }\href@noop {} {\emph {\bibinfo {title} {{The Electron-Phonon
  Interaction in Metals (Selected Topics in Solid State Physics XVI)}}}}\
  (\bibinfo  {publisher} {Elsevier Science Ltd},\ \bibinfo {year}
  {1981})\BibitemShut {NoStop}%
\bibitem [{\citenamefont {Lisowski}\ \emph {et~al.}(2004)\citenamefont
  {Lisowski}, \citenamefont {Loukakos}, \citenamefont {Bovensiepen},
  \citenamefont {St�hler}, \citenamefont {Gahl},\ and\ \citenamefont
  {Wolf}}]{Lisowski2004}%
  \BibitemOpen
  \bibfield  {author} {\bibinfo {author} {\bibfnamefont {M.}~\bibnamefont
  {Lisowski}}, \bibinfo {author} {\bibfnamefont {P.}~\bibnamefont {Loukakos}},
  \bibinfo {author} {\bibfnamefont {U.}~\bibnamefont {Bovensiepen}}, \bibinfo
  {author} {\bibfnamefont {J.}~\bibnamefont {St�hler}}, \bibinfo {author}
  {\bibfnamefont {C.}~\bibnamefont {Gahl}}, \ and\ \bibinfo {author}
  {\bibfnamefont {M.}~\bibnamefont {Wolf}},\ }\href {\doibase
  10.1007/s00339-003-2301-7} {\bibfield  {journal} {\bibinfo  {journal}
  {Applied Physics A: Materials Science \& Processing}\ }\textbf {\bibinfo
  {volume} {78}},\ \bibinfo {pages} {165} (\bibinfo {year} {2004})}\BibitemShut
  {NoStop}%
\bibitem [{\citenamefont {Bovensiepen}(2007)}]{Bovensiepen2007}%
  \BibitemOpen
  \bibfield  {author} {\bibinfo {author} {\bibfnamefont {U.}~\bibnamefont
  {Bovensiepen}},\ }\href {\doibase 10.1088/0953-8984/19/8/083201} {\bibfield
  {journal} {\bibinfo  {journal} {Journal of Physics: Condensed Matter}\
  }\textbf {\bibinfo {volume} {19}},\ \bibinfo {pages} {083201} (\bibinfo
  {year} {2007})}\BibitemShut {NoStop}%
\bibitem [{\citenamefont {Cheng}\ \emph {et~al.}(2010)\citenamefont {Cheng},
  \citenamefont {Song}, \citenamefont {Zhang}, \citenamefont {Zhang},
  \citenamefont {Wang}, \citenamefont {Jia}, \citenamefont {Wang},
  \citenamefont {Wang}, \citenamefont {Zhu}, \citenamefont {Chen},
  \citenamefont {Ma}, \citenamefont {He}, \citenamefont {Wang}, \citenamefont
  {Dai}, \citenamefont {Fang}, \citenamefont {Xie}, \citenamefont {Qi},
  \citenamefont {Liu}, \citenamefont {Zhang},\ and\ \citenamefont
  {Xue}}]{Cheng2010}%
  \BibitemOpen
  \bibfield  {author} {\bibinfo {author} {\bibfnamefont {P.}~\bibnamefont
  {Cheng}}, \bibinfo {author} {\bibfnamefont {C.}~\bibnamefont {Song}},
  \bibinfo {author} {\bibfnamefont {T.}~\bibnamefont {Zhang}}, \bibinfo
  {author} {\bibfnamefont {Y.}~\bibnamefont {Zhang}}, \bibinfo {author}
  {\bibfnamefont {Y.}~\bibnamefont {Wang}}, \bibinfo {author} {\bibfnamefont
  {J.-F.}\ \bibnamefont {Jia}}, \bibinfo {author} {\bibfnamefont
  {J.}~\bibnamefont {Wang}}, \bibinfo {author} {\bibfnamefont {Y.}~\bibnamefont
  {Wang}}, \bibinfo {author} {\bibfnamefont {B.-F.}\ \bibnamefont {Zhu}},
  \bibinfo {author} {\bibfnamefont {X.}~\bibnamefont {Chen}}, \bibinfo {author}
  {\bibfnamefont {X.}~\bibnamefont {Ma}}, \bibinfo {author} {\bibfnamefont
  {K.}~\bibnamefont {He}}, \bibinfo {author} {\bibfnamefont {L.}~\bibnamefont
  {Wang}}, \bibinfo {author} {\bibfnamefont {X.}~\bibnamefont {Dai}}, \bibinfo
  {author} {\bibfnamefont {Z.}~\bibnamefont {Fang}}, \bibinfo {author}
  {\bibfnamefont {X.}~\bibnamefont {Xie}}, \bibinfo {author} {\bibfnamefont
  {X.-L.}\ \bibnamefont {Qi}}, \bibinfo {author} {\bibfnamefont {C.-X.}\
  \bibnamefont {Liu}}, \bibinfo {author} {\bibfnamefont {S.-C.}\ \bibnamefont
  {Zhang}}, \ and\ \bibinfo {author} {\bibfnamefont {Q.-K.}\ \bibnamefont
  {Xue}},\ }\href {\doibase 10.1103/PhysRevLett.105.076801} {\bibfield
  {journal} {\bibinfo  {journal} {Physical Review Letters}\ }\textbf {\bibinfo
  {volume} {105}},\ \bibinfo {pages} {076801} (\bibinfo {year}
  {2010})}\BibitemShut {NoStop}%
\bibitem [{\citenamefont {Rauh}\ \emph {et~al.}(1981)\citenamefont {Rauh},
  \citenamefont {Geick}, \citenamefont {Kohler}, \citenamefont {Nucker},\ and\
  \citenamefont {Lehner}}]{Rauh1981}%
  \BibitemOpen
  \bibfield  {author} {\bibinfo {author} {\bibfnamefont {H.}~\bibnamefont
  {Rauh}}, \bibinfo {author} {\bibfnamefont {R.}~\bibnamefont {Geick}},
  \bibinfo {author} {\bibfnamefont {H.}~\bibnamefont {Kohler}}, \bibinfo
  {author} {\bibfnamefont {N.}~\bibnamefont {Nucker}}, \ and\ \bibinfo {author}
  {\bibfnamefont {N.}~\bibnamefont {Lehner}},\ }\href {\doibase
  10.1088/0022-3719/14/20/009} {\bibfield  {journal} {\bibinfo  {journal}
  {Journal of Physics C: Solid State Physics}\ }\textbf {\bibinfo {volume}
  {14}},\ \bibinfo {pages} {2705} (\bibinfo {year} {1981})}\BibitemShut
  {NoStop}%
\bibitem [{\citenamefont {Hashimoto}\ \emph {et~al.}(2010)\citenamefont
  {Hashimoto}, \citenamefont {He}, \citenamefont {Tanaka}, \citenamefont
  {Testaud}, \citenamefont {Meevasana}, \citenamefont {Moore}, \citenamefont
  {Lu}, \citenamefont {Yao}, \citenamefont {Yoshida}, \citenamefont {Eisaki},
  \citenamefont {Devereaux}, \citenamefont {Hussain},\ and\ \citenamefont
  {Shen}}]{Hashimoto2010}%
  \BibitemOpen
  \bibfield  {author} {\bibinfo {author} {\bibfnamefont {M.}~\bibnamefont
  {Hashimoto}}, \bibinfo {author} {\bibfnamefont {R.-H.}\ \bibnamefont {He}},
  \bibinfo {author} {\bibfnamefont {K.}~\bibnamefont {Tanaka}}, \bibinfo
  {author} {\bibfnamefont {J.-P.}\ \bibnamefont {Testaud}}, \bibinfo {author}
  {\bibfnamefont {W.}~\bibnamefont {Meevasana}}, \bibinfo {author}
  {\bibfnamefont {R.~G.}\ \bibnamefont {Moore}}, \bibinfo {author}
  {\bibfnamefont {D.}~\bibnamefont {Lu}}, \bibinfo {author} {\bibfnamefont
  {H.}~\bibnamefont {Yao}}, \bibinfo {author} {\bibfnamefont {Y.}~\bibnamefont
  {Yoshida}}, \bibinfo {author} {\bibfnamefont {H.}~\bibnamefont {Eisaki}},
  \bibinfo {author} {\bibfnamefont {T.~P.}\ \bibnamefont {Devereaux}}, \bibinfo
  {author} {\bibfnamefont {Z.}~\bibnamefont {Hussain}}, \ and\ \bibinfo
  {author} {\bibfnamefont {Z.-X.}\ \bibnamefont {Shen}},\ }\href {\doibase
  10.1038/nphys1632} {\bibfield  {journal} {\bibinfo  {journal} {Nature
  Physics}\ }\textbf {\bibinfo {volume} {6}},\ \bibinfo {pages} {414} (\bibinfo
  {year} {2010})}\BibitemShut {NoStop}%
\bibitem [{\citenamefont {Moritz}\ \emph {et~al.}(2011)\citenamefont {Moritz},
  \citenamefont {Johnston}, \citenamefont {Devereaux}, \citenamefont
  {Muschler}, \citenamefont {Prestel}, \citenamefont {Hackl}, \citenamefont
  {Lambacher}, \citenamefont {Erb}, \citenamefont {Komiya},\ and\ \citenamefont
  {Ando}}]{Moritz2011}%
  \BibitemOpen
  \bibfield  {author} {\bibinfo {author} {\bibfnamefont {B.}~\bibnamefont
  {Moritz}}, \bibinfo {author} {\bibfnamefont {S.}~\bibnamefont {Johnston}},
  \bibinfo {author} {\bibfnamefont {T.~P.}\ \bibnamefont {Devereaux}}, \bibinfo
  {author} {\bibfnamefont {B.}~\bibnamefont {Muschler}}, \bibinfo {author}
  {\bibfnamefont {W.}~\bibnamefont {Prestel}}, \bibinfo {author} {\bibfnamefont
  {R.}~\bibnamefont {Hackl}}, \bibinfo {author} {\bibfnamefont
  {M.}~\bibnamefont {Lambacher}}, \bibinfo {author} {\bibfnamefont
  {A.}~\bibnamefont {Erb}}, \bibinfo {author} {\bibfnamefont {S.}~\bibnamefont
  {Komiya}}, \ and\ \bibinfo {author} {\bibfnamefont {Y.}~\bibnamefont
  {Ando}},\ }\href {\doibase 10.1103/PhysRevB.84.235114} {\bibfield  {journal}
  {\bibinfo  {journal} {Physical Review B}\ }\textbf {\bibinfo {volume} {84}},\
  \bibinfo {pages} {235114} (\bibinfo {year} {2011})}\BibitemShut {NoStop}%
\bibitem [{\citenamefont {McIver}\ \emph {et~al.}(2012)\citenamefont {McIver},
  \citenamefont {Hsieh}, \citenamefont {Steinberg}, \citenamefont
  {Jarillo-Herrero},\ and\ \citenamefont {Gedik}}]{McIver2012}%
  \BibitemOpen
  \bibfield  {author} {\bibinfo {author} {\bibfnamefont {J.~W.}\ \bibnamefont
  {McIver}}, \bibinfo {author} {\bibfnamefont {D.}~\bibnamefont {Hsieh}},
  \bibinfo {author} {\bibfnamefont {H.}~\bibnamefont {Steinberg}}, \bibinfo
  {author} {\bibfnamefont {P.}~\bibnamefont {Jarillo-Herrero}}, \ and\ \bibinfo
  {author} {\bibfnamefont {N.}~\bibnamefont {Gedik}},\ }\href {\doibase
  10.1038/nnano.2011.214} {\bibfield  {journal} {\bibinfo  {journal} {Nature
  nanotechnology}\ }\textbf {\bibinfo {volume} {7}},\ \bibinfo {pages} {96}
  (\bibinfo {year} {2012})}\BibitemShut {NoStop}%
\end{thebibliography}%

\end{document}